\documentclass[journal]{IEEEtran}

\usepackage{cite}
\usepackage{dblfloatfix}
\usepackage{mathtools}
\usepackage{float}
\usepackage{amsmath,amssymb,amsfonts}
\usepackage{algorithm}
\usepackage[noend]{algpseudocode}
\usepackage{graphicx}
\graphicspath{{./Figures/}}
\usepackage{textcomp}
\usepackage{comment}
\usepackage{gensymb} %for degrees symbol
\usepackage{booktabs}
\usepackage{siunitx}
\usepackage{bm} % bold and italic at once inside eqs
\usepackage{cite} %multi cite
\usepackage{cleveref} %multi ref
\usepackage{url}

\usepackage{array,multirow}
\usepackage[table]{xcolor}
\usepackage{xcolor,colortbl}
\def\BibTeX{{\rm B\kern-.05em{\sc i\kern-.025em b}\kern-.08em
T\kern-.1667em\lower.7ex\hbox{E}\kern-.125emX}}

\makeatletter
\renewcommand*\env@matrix[1][\arraystretch]{%
  \edef\arraystretch{#1}%
  \hskip -\arraycolsep
  \let\@ifnextchar\new@ifnextchar
  \array{*\c@MaxMatrixCols c}}
\makeatother

\begin{document}

%\title{An Optimal Control Strategy for Operating a Hybrid PV Plant to Provide Robust Power Reserves}

\title{Optimal Control Design for Operating a Hybrid PV Plant with Robust Power Reserves for Fast Frequency Regulation Services}


\author{Victor~Paduani,~\IEEEmembership{Member,~IEEE,}
        Qi~Xiao,~\IEEEmembership{Student Member,~IEEE,}
        Bei~Xu,~\IEEEmembership{Student Member,~IEEE,}
        David~Lubkeman,~\IEEEmembership{Fellow,~IEEE,}
        and~Ning~Lu,~\IEEEmembership{Fellow,~IEEE}% <-this % stops a space
\thanks{This research is supported by the U.S. Department of Energy's Office of Energy Efficiency and Renewable Energy (EERE) under the Solar Energy Technologies Office Award Number DE-EE0008770. 

V. Paduani is with the AGILe Lab, New York Power Authority, White Plains, NY 10601, USA (e-mail: victor.daldeganpaduani@nypa.gov). Q. Xiao, B. Xu, D. Lubkeman, and N. Lu are with the Department of Electrical and Computer Engineering, North Carolina State University, Raleigh, NC 27606, USA.}% <-this % stops a space
%\thanks{J. Doe and J. Doe are with Anonymous University.}% <-this % stops a space (e-mail: qxiao3@ncsu.edu, bxu8@ncsu.edu, dllubkem@ncsu.edu, nlu2@ncsu.edu)
%\thanks{Manuscript received May xx, 2022; revised June xx, 2022.}
}

\maketitle

\begin{abstract}
This paper presents an optimal control strategy for operating a solar hybrid system consisting of solar photovoltaic (PV) and a high-power, low-storage battery energy storage system (BESS). A state-space model of the hybrid PV plant is first derived, based on which an adaptive model predictive controller is designed. The controller's objective is to control the PV and BESS to follow  power setpoints sent to the the hybrid system while maintaining desired power reserves and meeting system operational contraints. Furthermore, an extended Kalman filter (EKF) is implemented for estimating the battery SOC, and an error sensitivity is executed to assess its limitations. To validate the proposed strategy, detailed EMT models of the hybrid system are developed so that losses and control limits can be quantified accurately. Day-long simulations are performed in an OPAL-RT real-time simulator using second-by-second actual PV farm data as inputs. Results verify that the proposed method can follow power setpoints while maintaining power reserves in days of high irradiance intermittency even with a small BESS storage.
\end{abstract}

%The inherent oscillatory action of P\&O-based FPPTs is leveraged to provide a sufficient measurement window for the real-time curve-fitting. Moreover, t


% \fi
\IEEEpeerreviewmaketitle

\section{Introduction}

\IEEEPARstart{T}{he} penetration of inverter-based resources (IBRs) is expected to increase sharply in many regional grids, and as conventional synchronous machines are substituted by IBRs, the grid's inertia will drastically reduce in the upcoming years. For instance, in \cite{yuan2020machine}, Yuan \emph{et al.} estimated a 1\% system inertia reduction for every 1\% increase in the photovoltaic (PV) penetration for the WECC system. This inertia decline leads to higher rate-of-change-of-frequency (ROCOF) and more severe frequency nadirs during contingencies, which may trigger under-frequency-load shedding and even the cascading tripping of generating units. To counter the aforementioned issues, new grid standards \cite{ieee1547ieee}\cite{california2016electric} have raised requirements for IBRs to provide grid support functions (GSFs), such as power curtailment, stronger disturbance ride-through characteristics, and response to high ROCOF events.

%incentives to expand the generation from renewable energy sources (RES) have significantly boosted the amount of inverter-based resources (IBRs) integrated into power systems. Consequently, as conventional synchronous machines are substitued by IBRs, the grid's inertia will drastically reduce in the upcoming years. For instance, in \cite{yuan2020machine}, Yuan \textit{et al. estimated a 1\% system inertia reduction for every 1\% increase in the photovoltaic (PV) penetration for the WECC system. Lowering the inertia increases the frequency nadirs during contingencies, which can trigger under frequency load shedding events, or even cause a cascaded tripping of units.

\begin{figure*}[!t]
	\centering
	\includegraphics[width=0.92\textwidth]{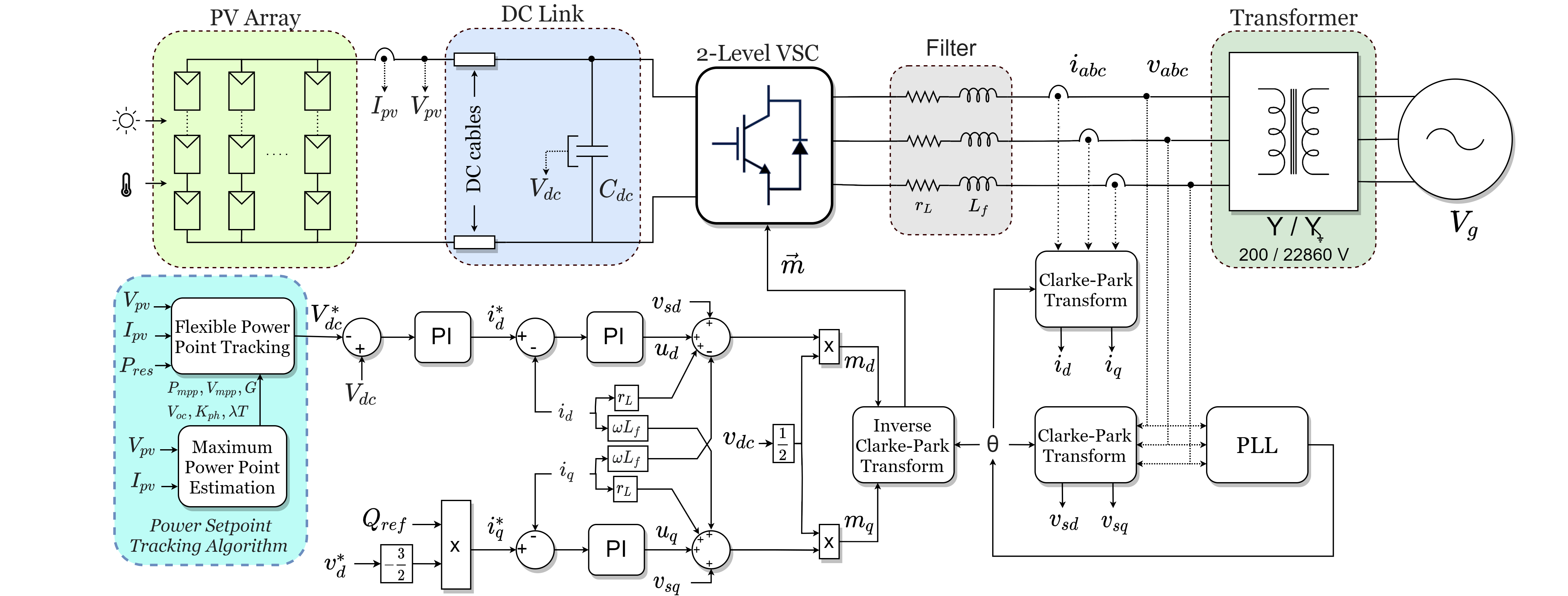}
	\caption{Circuit and control system block diagrams of a utility-scale PV system.}
	\label{maindiagram}
\end{figure*}

In \cite{paduani2021unified}, we introduced a PV power curtailment algorithm for maintaining accurate power reserves and providing fast frequency response (FFR) services to achieve higher frequency nadirs during contingencies. However, PV farms, operating by themselves, are still susceptible to fast clouding events. When irradiance drops sharply, even though a headroom is left for providing frequency support, the power reserves will quickly diminish, and consequently either the reserves or the power output will have to reduce. Therefore, using advanced power curtailment strategies alone cannot guarantee a PV plant to provide robust power reserves and follow power setpoints simultaneously under all weather conditions.

To provide robust power reserves, it is necessary for a PV plant to be coordinated with other resources. In \cite{chang2021coordinated}, Chang \textit{et al.} propose a coordination among multiple PV plants located in different geographic locations, so that if a shading event is imminent, power reserves are preemptively built up in other plants to help minimize the transient. Nevertheless, the approach is complicated, requiring coordination and communication among many PV plants, and by changing the PV injection in different regions, the system may experience voltage control issues due to PV injection variation. 

A promising approach is to form hybrid PV plants so that grid-following battery energy storage systems (BESS) can be dispatched for enhancing the capability of PV plants to follow power setpoints and/or maintain headrooms at all times. One advantage of this approach is that the BESS can also be used to blackstart the PV plant \cite{gevorgian2020provision} during blackouts.
%There are several methods in the literature that propose the coordination of BESS to address the intermittency from RES. 
In \cite{teleke2009control}, Teleke \emph{et al.} proposed the usage of state-of-charge (SOC) feedback strategy when integrating the BESS into a wind farm, whereas in \cite{daud2013improved}, Daud \emph{et al.} apply the same strategy to a PV plant without the power curtailment functionality. In \cite{teleke2010optimal}, Teleke \emph{et al.} improved their performance by substituting the SOC feedback strategy with optimal control techniques. 
Advantages of the optimal control approach (e.g., model predictive control (MPC)) include considering future changes in the control objective, accounting for inputs and outputs constraints, and robustness to modeling errors due to its inherent feedback control characteristic \cite{alaniz2004model}. 

Until now, an uncharted area is the control of intra-minute PV and BESS power outputs for meeting the reserve requirement and providing FFR.
In \cite{nair2021analysis}, Nair \emph{et al.} propose an MPC based algorithm for regulating PV and BESS in an energy scheduling problem with a step size of 5 minutes, which cannot deal with the intra-minute PV power fluctuations. In \cite{lei2017mpc}, Lei \emph{et al.} present the coordination of PV and BESS via optimal control in a smaller time scale, but the work is not focused on day-long operation and does not consider PV power reserves. BESS and PV intra-minute power coordination with detailed models is presented by Chen et al. in \cite{li2013battery}, which is a rule-based algorithm (not optimal) and does not account for PV curtailment or power reserves. Another drawback of the existing approaches is that most works use simplified BESS and/or PV models, which are insufficient for assessing the dynamic performance when providing fast grid services.  

Therefore, in this paper, we propose an optimal control strategy for operating a hybrid PV plant to provide reserves for regulation and FFR. The PV plant is equipped with the power curtailment algorithm recently introduced in \cite{paduani2021unified} for tracking power setpoints and maximum power point estimation (MPPE) under fast irradiance intermittency. A detailed lithium-ion battery model that has been experimentally validated in the literature is implemented for a more realistic approach, and an extended Kalman filter (EKF) is designed for estimating the battery SOC. Detailed EMT models for both the PV and the grid-following BESS units are developed in a real-time simulator from OPAL-RT. Day-long simulations with high resolution irradiance and temperature data collected by our industry partner, Strata Solar, are executed to analyze the capability of the hybrid PV plant to maintain power reserves while following a regulation D signal from PJM. 

The main contributions of the work to the literature are two-fold. \emph{First}, we develop an optimal control strategy for operating a hybrid PV plant that can maintain robust power reserves for regulation and/or FFR services. \emph{Second}, we derive a state-space (SS) model for the hybrid PV plant composed by an utility-scale PV system, a grid-following BESS unit, and an accurate second-order lithium-ion battery model. The proposed model is validated by day-long simulations using realistic data sets and detailed EMT models of the PV Plant and BESS. 
%\item Provided a systematic procedure for determining the service magnitude considering the PV and BESS sizes, battery lifetime depreciation, and reserves constraints.
    %\item \textcolor{red}{The main contributions of the work to the literature are summarized as:Provided a systematic procedure for determining the service magnitude considering the PV and BESS sizes, weather variability, battery lifetime depreciation, and reserve constraints.}
\section{Methodology}

\subsection{PV Plant Model}
\begin{figure}[htb]
	\centerline{\includegraphics[width=0.4\textwidth]{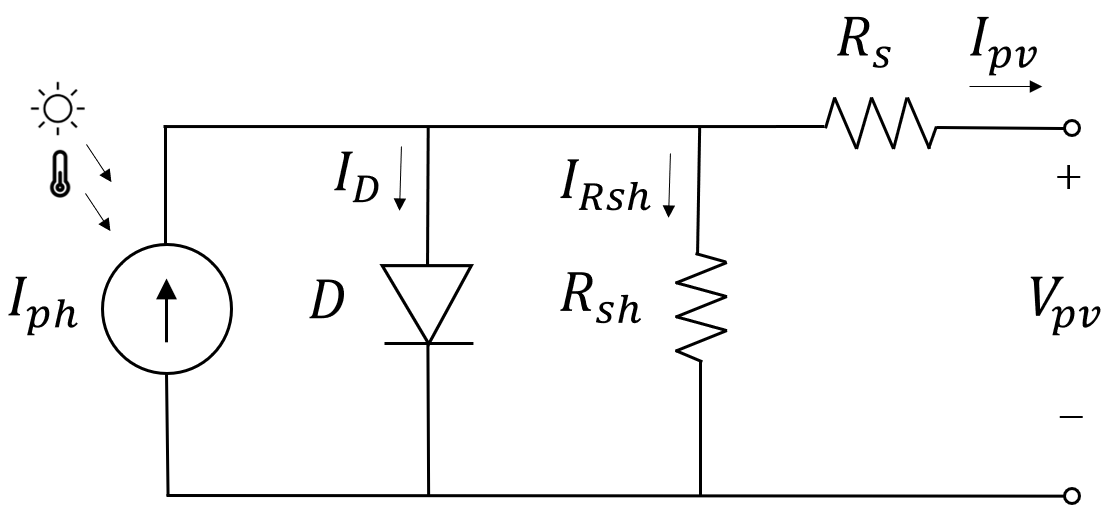}}
	\caption{Single-diode PV model circuit.}
	\label{Fig:pvcircuit}
\end{figure}

As shown in Fig. \ref{maindiagram}, the control system of the utility-scale PV plant uses a hierarchical control structure composed of a dc-link voltage controller cascaded with a current controller used to generate the inverter modulation signal, $\vec{m}$. More details on the converter controller can be found in \cite{paduani2021maximum}.
The PV array model shown in Fig. \ref{Fig:pvcircuit} is described by 
\begin{equation}
%\vspace{-0.5mm}
	I_{\mathrm{pv}}= I_{\mathrm{ph}} - I_{\mathrm{s}}\Big( e^{\frac{V_{\mathrm{pv}}+I_{\mathrm{pv}}R_{\mathrm{s}}}{a}}-1\Big) - \Big(\frac{V_{\mathrm{pv}}+I_{\mathrm{pv}}R_{\mathrm{s}}}{R_{\mathrm{sh}}}\Big)
	\label{PVeq}
\end{equation}
\begin{equation}
	V_{\mathrm{mp}} = \bigg(1+\frac{R_{\mathrm{s}}}{R_{\mathrm{sh}}}\bigg)a(w-1) - R_{\mathrm{s}}I_{\mathrm{ph}} \bigg(1-\frac{1}{w}\bigg)
	\label{eq:vmpp}
\end{equation}
\begin{equation}
	I_{\mathrm{mp}} = I_{\mathrm{ph}}\bigg(1-\frac{1}{w}\bigg) - \frac{a(w-1)}{R_{\mathrm{sh}}}
	\label{eq:impp}
\end{equation}
\begin{equation}
    P_{\mathrm{mp}} = V_{\mathrm{mp}}I_{\mathrm{mp}}
    \label{eq:Pmp2}
\end{equation}

\begin{equation}
	w = W\bigg\{ I_{\mathrm{ph}} \frac{e}{I_{\mathrm{s}}}\bigg\}
	\label{eq:w-Lambert}
\end{equation}
where $a$ is an ideality factor; $R_{\mathrm{s}}$ and $R_{\mathrm{sh}}$ are the series and shunt resistances, respectively; $I_{\mathrm{ph}}$ and $I_{\mathrm{s}}$ are the photo and diode saturation currents, respectively; and $w$, given by (\ref{eq:w-Lambert}), is calculated with the Lambert $W$ function.
The EMT models and the non-linear least squares Levenberg-Marquardt technique utilized for the MPPE are discussed in detail in \cite{paduani2021unified} and \cite{paduani2021implementation}.

%The five parameters of the PV array model are the ideality factor ($a$), the series resistance ($R_{\mathrm{s}}$), the shunt resistance ($R_{\mathrm{sh}}$), the photo current ($I_{\mathrm{ph}}$), and the diode saturation current ($I_{\mathrm{s}}$). The PV parameters are obtained in real-time using the non-linear least squares Levenberg-Marquardt technique presented in \cite{paduani2021unified} and \cite{paduani2021implementation}. Once the parameters are estimated, the voltage, current, and power at the MPP can be obtained by

%in which $w$, given by (\ref{eq:w-Lambert}), is calculated with the Lambert $W$ function.  

%A hierarchical control structure composed of a dc-link voltage controller cascaded with a current controller is used to generate the inverter modulation signal, $\vec{m}$. The five parameters of the model are the ideality factor ($a$), the series resistance ($R_{\mathrm{s}}$), the shunt resistance ($R_{\mathrm{sh}}$), the photocurrent ($I_{\mathrm{ph}}$), and the diode saturation current ($I_{\mathrm{s}}$).

%The five main PV parameters from Fig. \ref{Fig:pvcircuit} are obtained in real-time with the non-linear least squares Levenberg-Marquardt technique. The method for the real-time parameter estimation is validated and extensively discussed in \cite{paduani2021unified} and \cite{paduani2021implementation}; therefore, it is not discussed in this work. Once the parameters are estimated, the voltage, current, and power at the MPP can be obtained by 

\subsection{Grid-Following BESS Model}

\begin{figure}[!t]
	\centerline{\includegraphics[width=0.45\textwidth]{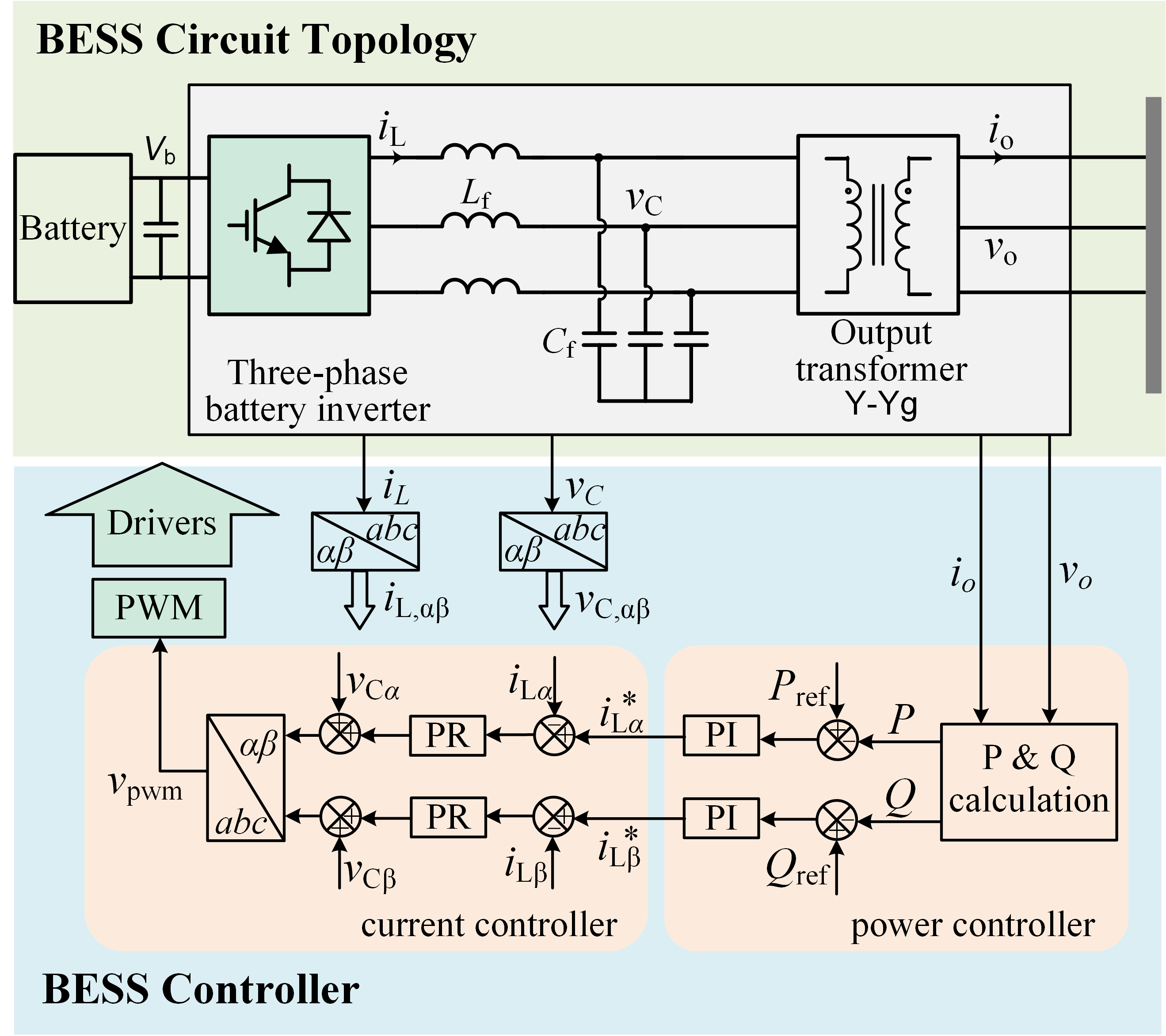}}
	\caption{BESS circuit and control system diagram.}
	\label{fig:BESS}
\end{figure}

The BESS unit, displayed in Fig. \ref{fig:BESS}, consists of (i) a lithium-ion battery, (ii) a three-phase inverter operating in grid-following mode, (iii) an LC output filter, and (iv) an output Y-Yg transformer. Figure \ref{fig:BESS} displays the BESS circuit and control system diagram. 

The control system is developed in the $\alpha\beta$ stationary reference frame (SRF), which requires one pair of PI controllers for power setpoint regulation, and one pair of proportional-resonant (PR) controllers for current regulation in the AC domain. The PR controllers transfer function is given by:

\begin{equation}
    G_{c} = K_{p} +\frac{2K_{r}\omega_{c}s}{s^{2} + 2\omega_{c}s + 4\omega_{0}^{2}}
    \label{eq:PR}
\end{equation}
where $K_{r}$ is the resonant gain at 2$\omega$, $\omega_{c}$ is the cut-off frequency, and $\omega_{0}$ is the grid's nominal frequency \cite{cha2009design}.

The modulation signal ($v_{\mathrm{pwm}}$) is generated by adding the output of the PR controllers with a feedforward signal of the output capacitor voltage. The authors present more details on the SRF control with an Y-Yg output transformer in \cite{xu2021novel}. The inverter is built with an averaged model of a two-level VSC developed in \cite{yazdani2010voltage}. Efficiency factors ($\eta_{\mathrm{charge}}$, $\eta_{\mathrm{discharge}}$) are used to represent the battery charging and discharging modes, respectively. %It is assumed that the BESS efficiency factors for charging and discharging modes are constant.

\subsection{Lithium-Ion Battery Model}

\begin{figure}[!t]
	\centerline{\includegraphics[width=0.5\textwidth]{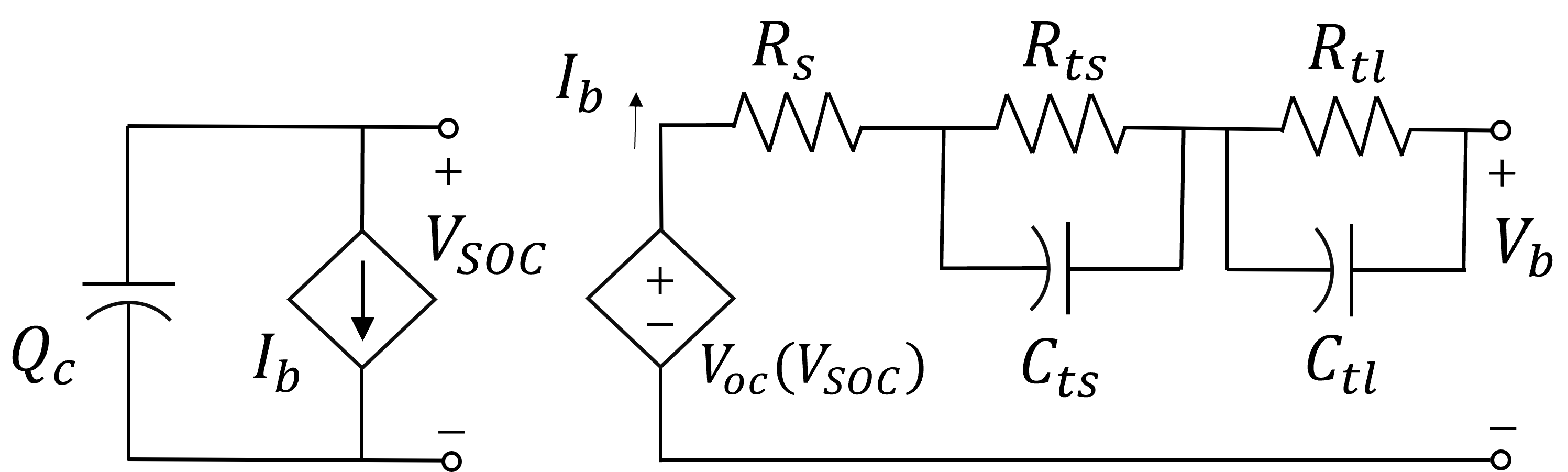}}
	\caption{Second-order equivalent lithium-ion battery model.}
	\label{fig:battery}
\end{figure}

Figure \ref{fig:battery} displays a second-order dynamic model of a lithium-ion battery that has been introduced in \cite{chen2006accurate}. The model is built by a combination of current and voltage sources that are used to recreate the realistic behavior of a lithium-ion battery under output current steps. The battery SOC is given by
\begin{equation}
    SOC(t) = SOC_{0} -  \int_{0}^{t} \frac{I_{\mathrm{b}}}{K_{\mathrm{age}}Q_{\mathrm{c}}} \,dt 
\end{equation}
where $Q_{\mathrm{c}}$ is the battery rated capacity and $K_{\mathrm{age}}$ is  an aging factor. In this paper, we consider real-time operation and set $K_{\mathrm{age}}=1$. 

The model contains two RC parallel branches ($R_{\mathrm{ts}}$, $C_{\mathrm{ts}}$) and ($R_{\mathrm{tl}}$, $C_{\mathrm{tl}}$) to represent the short-term and long-term voltage drops due to current step responses, respectively. Furthermore, a series resistor $R_{\mathrm{s}}$ is modeled to represent an internal instantaneous voltage drop due to the battery current. The dynamic response of the battery's internal voltages ($V_{\mathrm{Cts}}$, $V_{\mathrm{Ctl}}$) are given as follows.

\begin{equation}
    \frac{dV_{\mathrm{Cts}}}{dt} = \frac{I_{\mathrm{b}}}{C_{\mathrm{ts}}} - \frac{V_{\mathrm{Cts}}}{R_{\mathrm{ts}}C_{\mathrm{ts}}} 
\end{equation}

\begin{equation}
    \frac{dV_{\mathrm{Ctl}}}{dt} = \frac{I_{\mathrm{b}}}{C_{\mathrm{tl}}} - \frac{V_{\mathrm{Ctl}}}{R_{\mathrm{tl}}C_{\mathrm{tl}}} 
\end{equation}

The battery cell open-circuit voltage ($V_{\mathrm{oc}}$({\small $V_{\mathrm{SOC}}$})) is a function of the instantaneous SOC value ($V_{\mathrm{SOC}}$). Experiments can be carried to obtain the relation between the battery internal voltage and its SOC. In \cite{chen2012state}, the relation is obtained via curve-fitting, and a seventh-order polynomial is built as follows:
\begin{align}
\notag
    V_{oc}(V_{SOC}) = a_{1}{V_{SOC}}^{7} + a_{2}{V_{SOC}}^{6} + a_{3}{V_{SOC}}^{5} + a_{4}{V_{SOC}}^{4} + \\  a_{5}{V_{SOC}}^{3} + a_{6}{V_{SOC}}^{2} + 
    a_{7}V_{SOC} + a_{8}
    \label{eq:Voc}
\end{align}

Hence, the battery output voltage ($V_{\mathrm{b}}$) is defined by 
\begin{equation}
    V_{\mathrm{b}} = V_{\mathrm{oc}}(V_{\mathrm{SOC}}) - V_{\mathrm{Cts}} - V_{\mathrm{Ctl}} - R_{\mathrm{s}}I_{\mathrm{b}}
    \label{eq:Vb}
\end{equation}

It is important to mention that the RC parameters of the second-order lithium-ion battery model from \cite{chen2006accurate} are approximately constant between 20\% to 100\% SOC operation, but they change exponentially between 0\% to 20\% SOC. Consequently, the modeling approach presented here is only valid if the SOC is maintained above 20\% at all times. Moreover, a self-discharging resistor could be added in parallel to $Q_{\mathrm{c}}$ in Fig. \ref{fig:battery} to represent the battery self-discharge of 2-10\% per month, but this can be neglected for systems that are cycled often and/or do not leave the battery stored for a long time. Furthermore, as pointed in \cite{chen2006accurate}, in reality, all parameters from the second-order model are multivariable functions of SOC, current, temperature, and number of cycles. Nevertheless, the model can still present satisfactory performance for most application if a certain error tolerance is acceptable. Because the main focus of this work is not on developing an extremely accurate battery model, it is assumed the parameters are constant within the 20 to 100\% SOC range.

\subsection{Hybrid PV Plant State-Space Model}

The hybrid PV plant SS proposed in this work is designed with five state variables, which are assigned as: the battery SOC ($x_{1}$), the battery short-term internal voltage drop $V_{\mathrm{Cts}}$ ($x_{2}$), the battery long-term internal voltage drop $V_{\mathrm{Ctl}}$ ($x_{3}$), the battery current $I_{\mathrm{b}}$ ($x_{4}$), and the PV dc output power $P_{\mathrm{pv}}$ ($x_{5}$). Note the battery current and the PV output power are set as state variables so that constraints can be established for them in the MPC solver utilized in this work.%, which is required to ensure their physical limitations are respected during operation. 

The hybrid PV plant has two main control inputs: change in battery output current ($u_{1}$) and change in PV dc output power ($u_{2}$). The control inputs are set as derivatives instead of the actual battery current and PV output power so that state variables can be assigned to the actual battery current and PV output power. Furthermore, the system also contains one input disturbance, corresponding to the maximum available PV dc power, $P_{\mathrm{mp}}$ ($u_{3}$). The disturbance is a measurable signal that comes from the MPPE algorithm presented in \cite{paduani2021unified}. It is used to provide the MPC with constraints for the maximum PV power available. Inputs $u_{1}$ and $u_{2}$ are constrained within the ramp-up and ramp-down rate limits of the BESS and PV system by (\ref{eq:Ib_rate}) and (\ref{eq:Ppv_rate}). %It is assumed that the same ramping limit is implemented for both upwards and downwards power setpoint changes.
\begin{equation}
    u_{1} = \frac{dI_{\mathrm{b}}}{dt}, \quad -I_{\mathrm{b},\mathrm{rate}} \leq u_{1} \leq I_{\mathrm{b},\mathrm{rate}}
    \label{eq:Ib_rate}
\end{equation}
\begin{equation}
    u_{2} = \frac{dP_{\mathrm{pv}}}{dt}, \quad -P_{\mathrm{pv},\mathrm{rate}} \leq u_{2} \leq P_{\mathrm{pv},\mathrm{rate}}
    \label{eq:Ppv_rate}
\end{equation}
\begin{equation}
    u_{3} = P_{\mathrm{\mathrm{mp}}}, \quad \text{from (\ref{eq:Pmp2})}
\end{equation}

The model is designed with five output variables, which are assigned as: the hybrid plant output power ($y_{1}$), the battery current ($y_{2}$), the battery SOC ($y_{3}$), the power reserves ($y_{4}$), and the PV dc output power ($y_{5}$). The battery current, SOC, and the PV output power are defined as output variables to allow the implementation of soft constraints to the respective state variables when designing the MPC. Thus, the hybrid PV plant output power is given by (\ref{eq:Pout_y}), whereas the power reserves are given by (\ref{eq:Pres_y}).
\begin{equation}
    y_{1} = P_{\mathrm{out}} = \eta_{\mathrm{bess}}\Big[V_{oc}(x_{1}) - V_{\mathrm{ts}} - V_{\mathrm{tl}} - R_{\mathrm{s}}I_{\mathrm{b}}\Big]I_{\mathrm{b}} + \eta_{\mathrm{pv}}P_{\mathrm{pv}}
    \label{eq:Pout_y}
\end{equation}
\begin{align}
\notag
    y_{4} = P_{\mathrm{res}} = \eta_{\mathrm{bess}}\Big[P_{\mathrm{bess}}^{N} -\big(V_{oc}(x_{1}) - V_{\mathrm{ts}} - V_{\mathrm{tl}} - R_{\mathrm{s}}I_{\mathrm{b}}\big)I_{\mathrm{b}}\Big] + \\
   \eta_{\mathrm{pv}}(P_{\mathrm{mp}} - P_{\mathrm{pv}})
   \label{eq:Pres_y}
\end{align}
In (\ref{eq:Pres_y}), $P_{\mathrm{bess}}^{N}$ is the battery nominal power. The PV and BESS unit efficiency factors ($\eta_{\mathrm{pv}}$, $\eta_{\mathrm{bess}}$) both include the inverter and output transformer losses. In addition, the BESS efficiency factor is based on the charging or discharging operation of the battery by (\ref{eq:bess_effc}). %Note that since the upward power regulation is the main challenge for maintaining power reserves (since PV can always be curtailed), 
%%%
%LATER MENTION ABOUT effc_bess*P_nominal
%%%
\begin{align}
    \eta_{\mathrm{bess}} = 
\begin{cases}
    \eta_{\mathrm{discharge}} & \text{if $I_{\mathrm{b}} \geq 0$} \\
    \eta_{\mathrm{charge}} & \text{if $I_{\mathrm{b}} < 0$}
\end{cases}
\label{eq:bess_effc}
\end{align}

Thus, the SS of the system is defined as follows.
\begin{align}
    \notag
    \dot{\mathbf{x}} = A\mathbf{x} + B\mathbf{u}  \\
    \mathbf{y} = C\mathbf{x} + D\mathbf{u}
\end{align}
\begin{equation}
    \begin{bmatrix}
        \dot{x_{1}} \\[0.1cm]
        \dot{x_{2}} \\[0.1cm]
        \dot{x_{3}} \\[0.1cm]
        \dot{x_{4}} \\[0.1cm]
        \dot{x_{5}}
    \end{bmatrix}
    = 
    \begingroup
    \setlength\arraycolsep{1.5pt}
    \begin{bmatrix}
        0 & 0 & 0 & \tfrac{\text{-}1}{Q_{\mathrm{c}}} & 0\\[0.1cm]
        0 & \tfrac{\text{-}1}{R_{\mathrm{ts}}C_{\mathrm{ts}}} & 0 & \tfrac{1}{C_{\mathrm{ts}}} & 0\\[0.1cm]
        0 & 0 & \tfrac{\text{-}1}{R_{\mathrm{tl}}C_{\mathrm{tl}}} & \tfrac{1}{C_{\mathrm{tl}}} & 0 \\[0.1cm]
        0 & 0 & 0 & 0 & 0 \\[0.1cm]
        0 & 0 & 0 & 0 & 0
    \end{bmatrix}   
    \endgroup
    \hspace{-1pt}
    \begin{bmatrix}
        x_{1} \\[0.1cm]
        x_{2} \\[0.1cm]
        x_{3} \\[0.1cm]
        x_{4} \\[0.1cm]
        x_{5}
    \end{bmatrix}
    +
    \begingroup
    \setlength\arraycolsep{1.5pt}
    \begin{bmatrix}
        0 & 0 & 0 \\[0.1cm]
        0 & 0 & 0 \\[0.1cm]
        0 & 0 & 0 \\[0.1cm]
        1 & 0 & 0 \\[0.1cm]
        0 & 1 & 0 \\
    \end{bmatrix}
    \endgroup
    \begin{bmatrix}
        u_{1}\\
        u_{2}\\
        u_{3}
    \end{bmatrix}
    \label{eq:A_matrix}
\end{equation}

\begin{align}
\notag
    \begin{bmatrix}
        y_{1} \\[0.1cm]
        y_{2} \\[0.1cm]
        y_{3} \\[0.1cm]
        y_{4} \\[0.1cm]
        y_{5}
    \end{bmatrix}
    = 
    \begingroup
    \setlength\arraycolsep{2pt}
    \begin{bmatrix}
        0 & 0 & 0 &  \eta_{\mathrm{bess}} \Big[V_{\mathrm{oc}}(x_{1}) \textrm{-} x_{2} \textrm{-} x_{3} \textrm{-} R_{s}x_{4}\Big] & \eta_{\mathrm{pv}}\\[0cm]
        0 & 0 & 0 & 1 & 0\\[0cm]
        1 & 0 & 0 & 0 & 0\\[0cm]
        0 & 0 & 0 & \textrm{-}\eta_{\mathrm{bess}} \Big[V_{\mathrm{oc}}(x_{1}) \textrm{-} x_{2} \textrm{-} x_{3} \textrm{-} R_{s}x_{4}\Big] & \textrm{-}\eta_{\mathrm{pv}} \\[0cm]
        0 & 0 & 0 & 0 & 1
    \end{bmatrix}    
    \endgroup
    \begin{bmatrix}
        x_{1} \\[0.1cm]
        x_{2} \\[0.1cm]
        x_{3} \\[0.1cm]
        x_{4} \\[0.1cm]
        x_{5}
    \end{bmatrix}
    +
    \\ 
    \begin{bmatrix}
        0 & 0 & 0 & 0 & 0\\[0cm]
        0 & 0 & 0 & 0 & 0\\[0cm]
        0 & 0 & 0 & \eta_{\mathrm{pv}} & 0
    \end{bmatrix}^{T}
    \begin{bmatrix}
        u_{1}\\
        u_{2}\\
        u_{3}
    \end{bmatrix}
    \textcolor{white}{+}
    \hspace*{1.5cm}
    \label{eq:C_matrix}
\end{align}

\noindent
The battery open-circuit voltage, $V_{\mathrm{oc}}(x_{1})$, is defined by (\ref{eq:Voc}). It is worth mentioning the initial condition of the power reserves is set as $\eta_{\mathrm{bess}}P_{\mathrm{bess}}^{N}$.  Furthermore, note the controllability matrix of the proposed SS model, given by R = [B \hspace{0.25em} AB \hspace{0.25em} A$^{2}$B \hspace{0.25em} ... \hspace{0.25em} A$^{n-1}$B], presents full rank, confirming that all states of the system are controllable.

\subsection{Model Predictive Control Design}

\begin{figure}[!htb]
	\centerline{\includegraphics[width=0.5\textwidth]{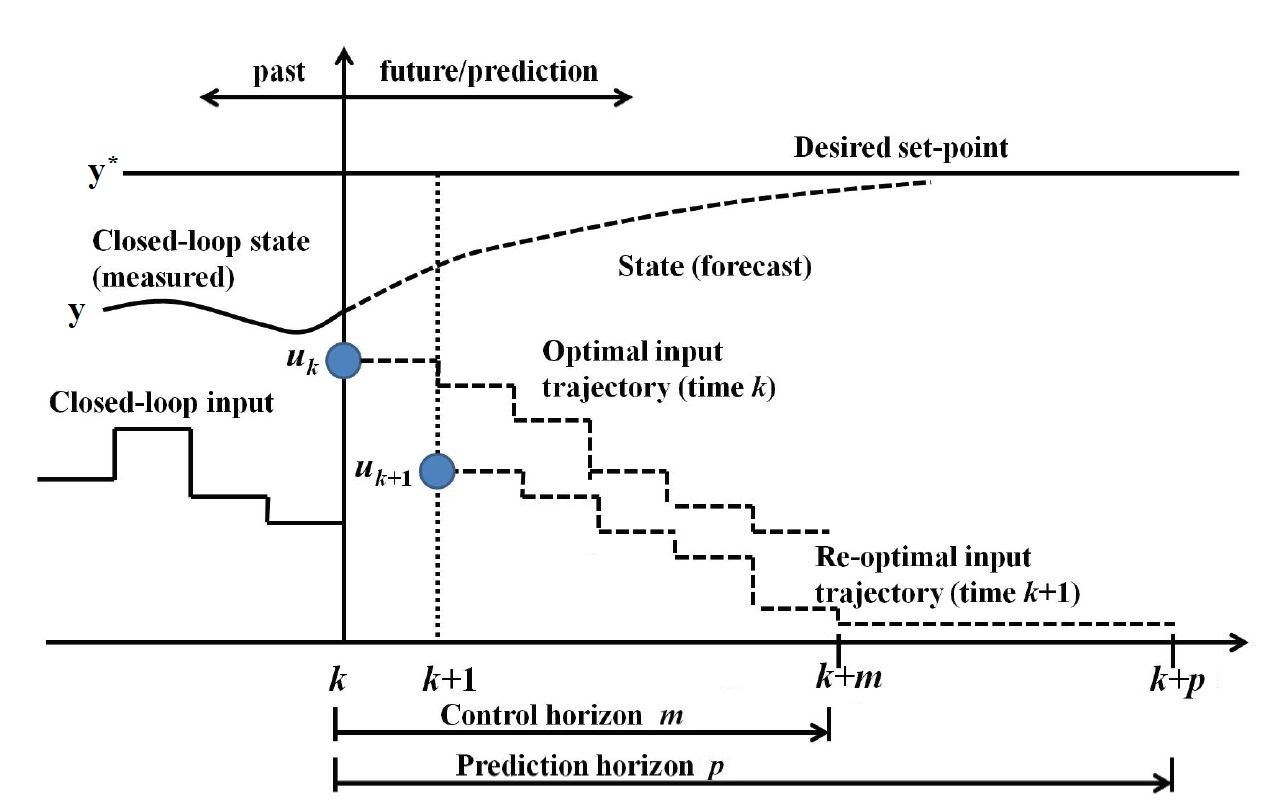}}
	\caption{MPC prediction and control horizons. Modified from \cite{dai2012discrete}.}
	\label{fig:MPC_horizon}
\end{figure}

In this work, the optimal operation of the hybrid PV plant is achieved by the implementation of a MPC, which utilizes feedback control to solve for the optimal operation of the system while maintaining inputs and outputs within constraints that respect the physical limitations of the components. Figure \ref{fig:MPC_horizon} depicts the operation of a MPC. The prediction horizon, $p$, at time $k$, represents how many steps ahead the controller considers for the optimization, whereas the control horizon, $m$, at time $k$, corresponds to how many control steps are optimized within a given prediction horizon. Note that when the MPC is designed with $m<p$, the optimal inputs, $u_k$, are fixed at the last optimized value of the control horizon until the end of the prediction horizon, $p$. 

The objective of the optimal control problem is to find a sequence of optimal control signals, $z_{k}$, given in (\ref{eq:zk}), to follow the targeted output signals, $y_{j}(i)$, $i\in p$.  
\begin{align}
\notag
    z_{k}^{T} = \Big[ u(k)^{T} \hspace{0.5em} u(k+1|k)^{T} \hspace{0.5em} ... \hspace{0.5em} u(k + p -1|k)^{T}   \Big],\\
    \text{if }m < p \quad u(k+m+1:k+p-1|k)^{T} = u(k+m|k)^{T}
    \label{eq:zk}
\end{align}

The MPC problem can be formulated as 
\begin{equation}
   \min J_{\epsilon}(z_{k}) + J_{y}(z_{k}) + J_{\Delta u}(z_{k})  
   %J(z_{k}) = J_{y}(z_{k}) + J_{\Delta u}(z_{k}) + J_{\epsilon}(z_{k})     \label{eq:Jcost}
\end{equation}
\begin{equation}
    J_{y}(z_{k}) = \sum_{j=1}^{n_{y}} \sum_{i=1}^{p} \Bigg\{ \frac{w_{j}^{y}}{s_{j}^{y}} \bigg[(y_{j}^{*}(k+i|k) - y_{j}(k+i|k) \bigg]\Bigg\} ^{2}
    \label{eq:Jy_cost}
\end{equation}
\begin{equation}
    J_{\Delta u}(z_{k}) = \sum_{j=1}^{n_{u}} \sum_{i=0}^{p-1} \Bigg\{ \frac{w_{j}^{\Delta u}}{s_{j}^{u}} \bigg[(u_{j}(k+i|k) - u_{j}(k+i -1|k) \bigg]\Bigg\} ^{2}
    \label{eq:J_delta}
\end{equation}
where $J_{y}(z_{k})$ is the tracking error cost, $J_{\Delta u}(z_{k})$ is the cost associated with penalizing aggressive control moves, $J_{\epsilon}$ is a penalty factor associated with constraint violations, $w_{j}^{y}$ and $s_{j}^{y}$ are the cost weights and scale factors for the jth plant output ($y_{j}$), respectively, $k$ is the current control interval, $p$ is the size of the prediction horizon, $n_{y}$ is the number of plant output variables, $w_{j}^{\Delta u}$ and $s_{j}^{u}$ are the cost weights and scale factors for the jth control input ($u_{j}$), respectively, and $n_{u}$ is the number of plant control signals. 

The constraint violation cost, $J_{\epsilon}(z_{k})$, ensures the solver can maintain numerical stability during real-time operation if the solution becomes unfeasible. Furthermore, to avoid numerical instabilities caused by hard constraints, the battery current, battery SOC, and the power reserves are set with soft constraints. More details on the penalty factor for handling constraint violations are presented in \cite{bemporad2010model}. 

In (\ref{eq:Jy_cost}), the tracking cost weights ($w_{j}^{y}$) must be tuned so the desired operation performance is achieved. For the hybrid PV plant, the main goal is to track an output power reference defined in (\ref{eq:Pout_y}), hence its weight is set with the highest value. Moreover, some of the weights will always be zero, such as the weights of the battery current ($w_{2}^{y}$) and the PV output power ($w_{5}^{y}$), because there is no specific goal for each of them individually.

In (\ref{eq:J_delta}), the penalization of aggressive control moves incentives the controller to maintain a smooth operation when trying to optimize the tracking error cost, and helps maintaining good numerical conditioning.

%The SOC weight ($w_{3}^{y}$) could be set as zero as well, because that will allow the MPC to optimize the operation without taking into account any given setpoint for the battery SOC. However, that approach has two drawbacks. First, by not having an SOC setpoint, the initial SOC value for each new operating cycle cannot be controlled, which can be a problem if a new cycle starts with the battery fully discharged. Second, during moments of high forecast error, if the battery is at low SOC levels, it may not be available to help boost the power output during PV drops, thus it would be preferred to maintain the SOC at higher levels when possible, considering that downward regulation can always be easily achieved by curtailing the PV.

The weight of the battery SOC ($w_{3}^{y}$) is set to a small value with a high setpoint (e.g., 90\%). With those settings, the MPC will boost the battery SOC to higher levels when possible, but it will significantly prioritize the power setpoint tracking during operation. It is worth mentioning that leaving a lithium-ion battery to linger at a fully charged state will accelerate their aging process \cite{vetter2005ageing}; therefore, it may be preferred to not set the BESS SOC setpoint to 100\%.

\begin{figure}[!htb]
	\centerline{\includegraphics[width=0.45\textwidth]{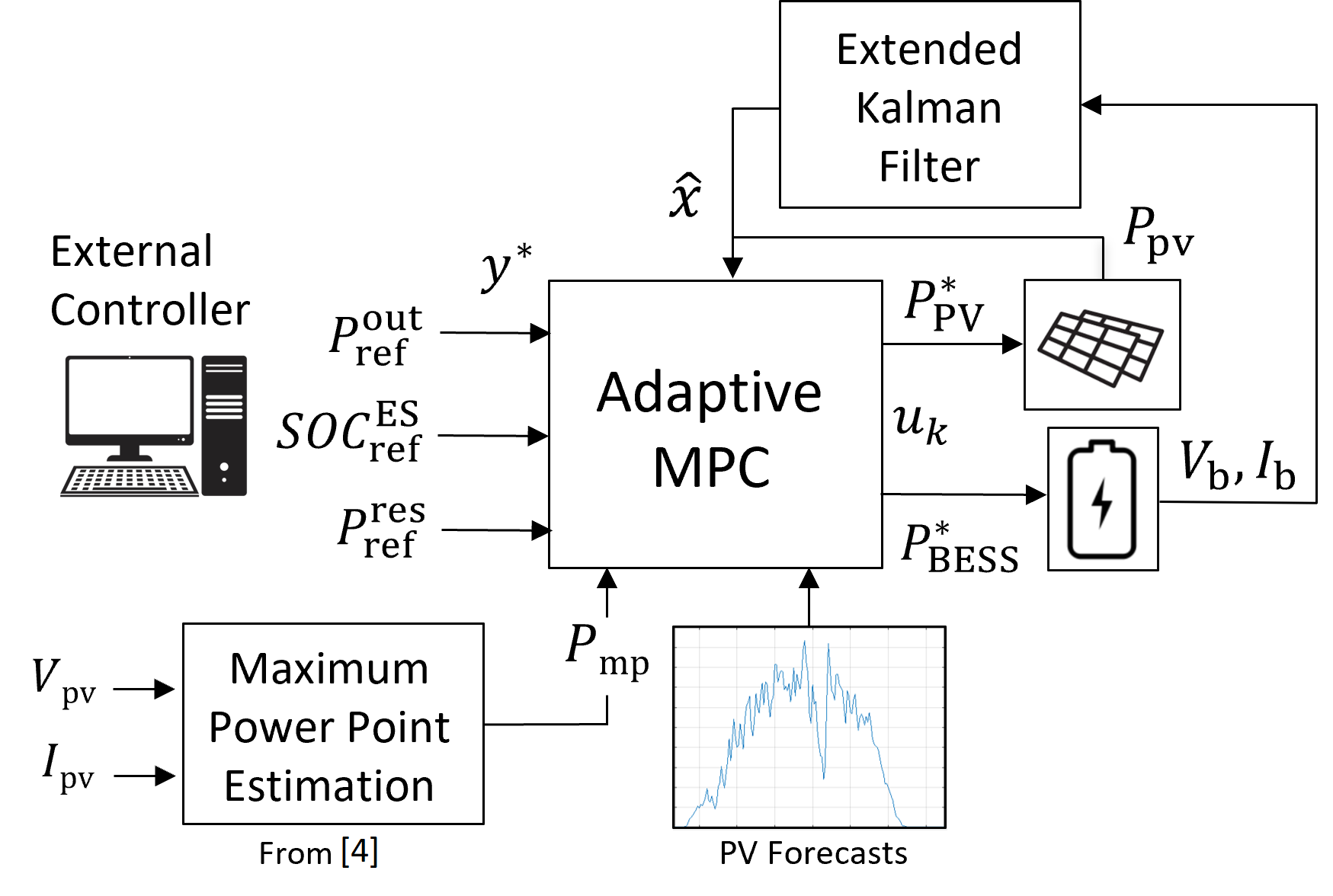}}
	\caption{Configuration of the model predictive controller.}
	\label{fig:MPC_diagram}
\end{figure}

Due to the nonlinear characteristics of the hybrid PV plant SS model, traditional linear MPC design is not appropriate. The MPC Toolbox from \cite{bemporad2010model} is compatible with the eMEGASIM environment from OPAL-RT utilized for real-time simulations and provides multiple options for the implementation of MPCs that can handle nonlinear models, such as: (a) Adaptive MPC, (b) Gain-Scheduled MPC, and (c) Nonlinear MPC. In this work, best performance was found with the adaptive MPC. Figure \ref{fig:MPC_diagram} displays the proposed control strategy. The adaptive MPC sends the power setpoints to the plant comprised of the PV and BESS unit. Sensors measure the PV output power as well as the battery voltage and current, which are inserted in an EKF that estimates the system states. PV forecasts and information about the instantaneous maximum available PV power are also inputs of the adaptive MPC. In the figure, the hybrid PV plant setpoints are generated by an external energy management system (EMS) running in an external controller. 

%During the development of this work, both the adaptive MPC and the nonlinear MPC were tested. However, it was soon observed that the computational burden from the nonlinear MPC would significantly limit the prediction horizon's length, forcing it to be up to nine times smaller than the prediction horizon of the adaptive MPC for a similar computational processing time. Because a significantly smaller prediction horizon would not be able to manage the BESS SOC due to its slow dynamics, the nonlinear MPC approach was discarded. 

Three main parameters must be defined when designing the MPC: timestep ($T_{\mathrm{mpc}}$), prediction horizon, and the control horizon. In this work, the timestep is set as 3 seconds, whereas the prediction horizon is selected as 400 steps (20 minutes), which is long enough to account for the battery slow charging/discharging dynamics. Note the timestep could be reduced to 2 or even 1 second(s) if real-time operation can be maintained by the real-time simulator, but it should not be increased further, as this will deteriorate the MPC's response to irradiance transients. The control horizon is set to 20 steps, based on the length of the ultra short-term PV forecasting. The estimated PV power ($P_{\mathrm{mp}}$) is used both as a disturbance applied to the power reserves and as an upper constraint to the PV output power. 

% The timestep should be small enough for the method to respond to external disturbances (irradiance changes) and properly respond to the system's dynamics, but it cannot be too small or the computational burden will be too high for real-time operation. Higher prediction and control horizons will also increase the computational effort required, and will not necessarily improve the system performance. For example, if the prediction horizon is too long, part of its results may be useless if unexpected disturbances occur.

By actuating as a time-varying upper constraint being updated in real time, $P_{\mathrm{mp}}$ forces the MPC to consider what is the actual available PV power when solving for the optimal control commands. It is worth mentioning that time-varying upper bounds capable of being updated in real time is a feature from the MPC Toolbox \cite{mpc_releasenotes} that only became compatible with RT-LAB version 2021.2, which is the platform used to run eMEGASIM models in real-time simulations.

%introduced into the MPC Toolbox on MATLAB R2020b version \cite{mpc_releasenotes}. Furthermore, RT-LAB only became compatible with MATLAB R2020b in its 2021.2 version, released on November 30, 2021. Consequently, the approach proposed here can only be carried out in RT-LAB versions 2021.2 or above. 
% Moreover, by applying the constraint to an output variable, the constraint can be established with some degree of softness, hence avoiding numerical instabilities that could occur when a hard constraint is violated. Admittedly, the small degree of softness may cause the MPC controller to command a few extra kWs from the PV plant when the irradiance oscillates; however, those small errors are insignificant considering that (i) if the command is larger than the available power, the PV will output the maximum power available, and (ii) modeling errors will always occur in practice.

\subsection{PV Forecast}

We consider the employment of two forecast techniques to the hybrid PV plant: (i) an ultra-short-term forecast of the available PV power with skycams \cite{schmidt2017short}, and a 20 minutes ahead forecast of the average available PV power.

It is assumed the 20-minutes ahead forecast performs with $\pm$10\% errors. For the ultra-short-term forecast, it is assumed that the PV plant runs the algorithm proposed in \cite{lipperheide2015embedded}, which has been experimentally validated with 70 inverters of a 48 MWdc plant. The method reported a relative root mean square error (rRMSE) of 3.2\% for its 20-seconds ahead forecast, and an approximately linear error increase up to 8.2\% at its 60-seconds ahead forecast. Therefore, we utilize the ultra-short-term forecasting only for the first 60 seconds, which is also defined as the length of the control horizon of the adaptive MPC. %By utilizing the forecast for the entire duration of the control horizon, the control trajectory optimization is based on realistic upper bound constraints for what the maximum available PV power will be. 

\begin{figure}[!htb]
	\centerline{\includegraphics[width=0.5\textwidth]{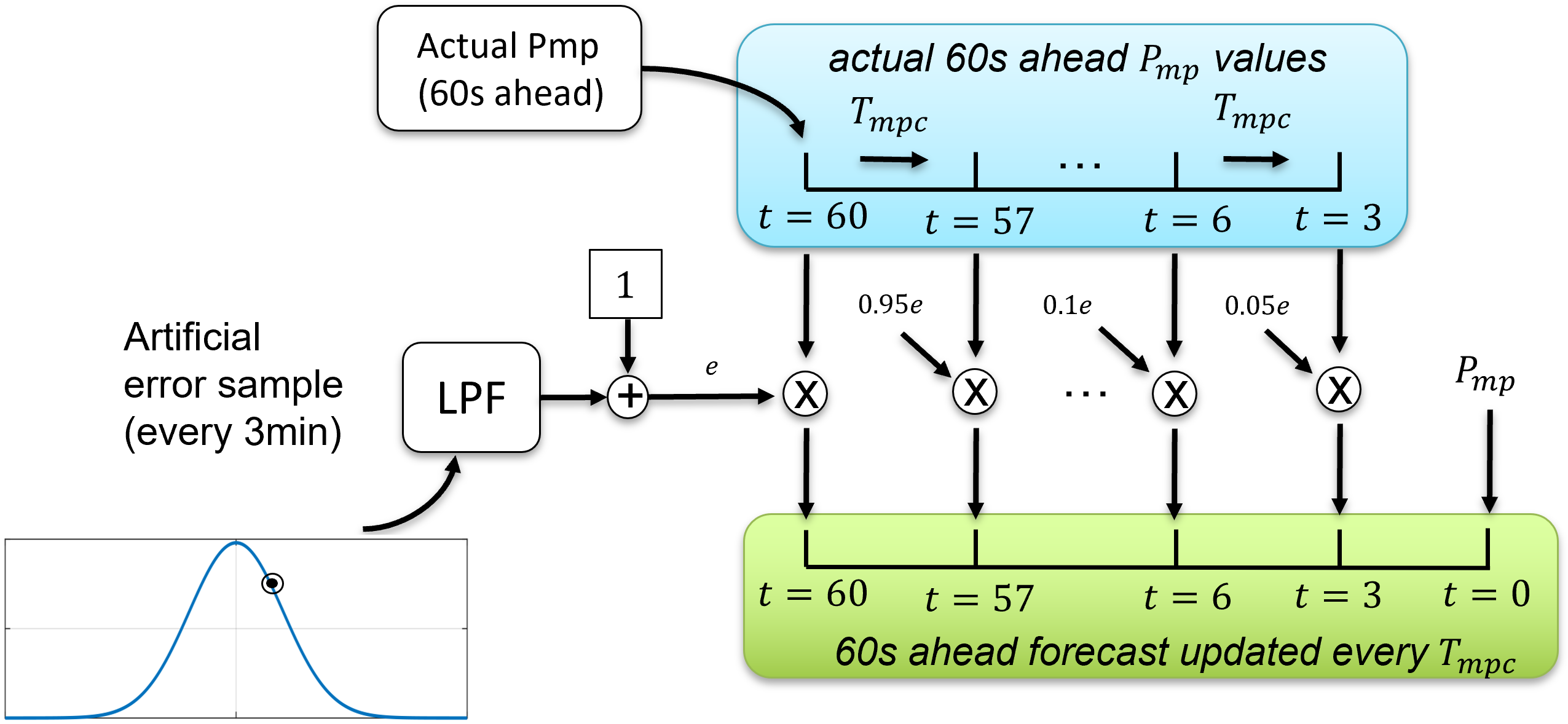}}
	\caption{Illustration of how the nowcasting performance is replicated.}
	\label{fig:nowcasting}
\end{figure}

Figure \ref{fig:nowcasting} demonstrates how the performance of the ultra-short-term forecasting method is simulated for a system with an MPC step of 3 seconds and a control horizon of 60 seconds. First, a normal distribution is sampled every few minutes to generate an artificial performance error. Next, the new error value is filtered by a low-pass filter with a time constant of one minute. The filter ensures the estimations will not be instantaneously stepped up or down as new error values are sampled, providing a more realistic transition between different estimation errors. Then, the filtered error factor ($e$) is applied to the actual $P_{\mathrm{mp}}$ values, which are constantly being stored and updated in an auxiliary vector. By reducing the error factor ($e$) from its full value at the 60-seconds ahead forecast down to $0.05e$ at the 3-seconds ahead forecast, the estimation accuracy is linearly improved as the prediction length is reduced. 

\subsection{Battery State Estimation}
In practice, the battery states cannot be directly measured. Thus, in this paper, we use an EKF (see Fig. \ref{fig:MPC_diagram}) to estimate the battery states ($\hat x$) using battery voltage and current measurements as inputs. Gaussian noises are added to the measurements to match the signal-to-noise (SNR) ratios expected from standard dc sensors. Because the EKF theory is well established in the literature, it will not be discussed here.

\section{Real-Time Simulation Results}

\begin{figure}[!htb]
	\centerline{\includegraphics[width=0.5\textwidth]{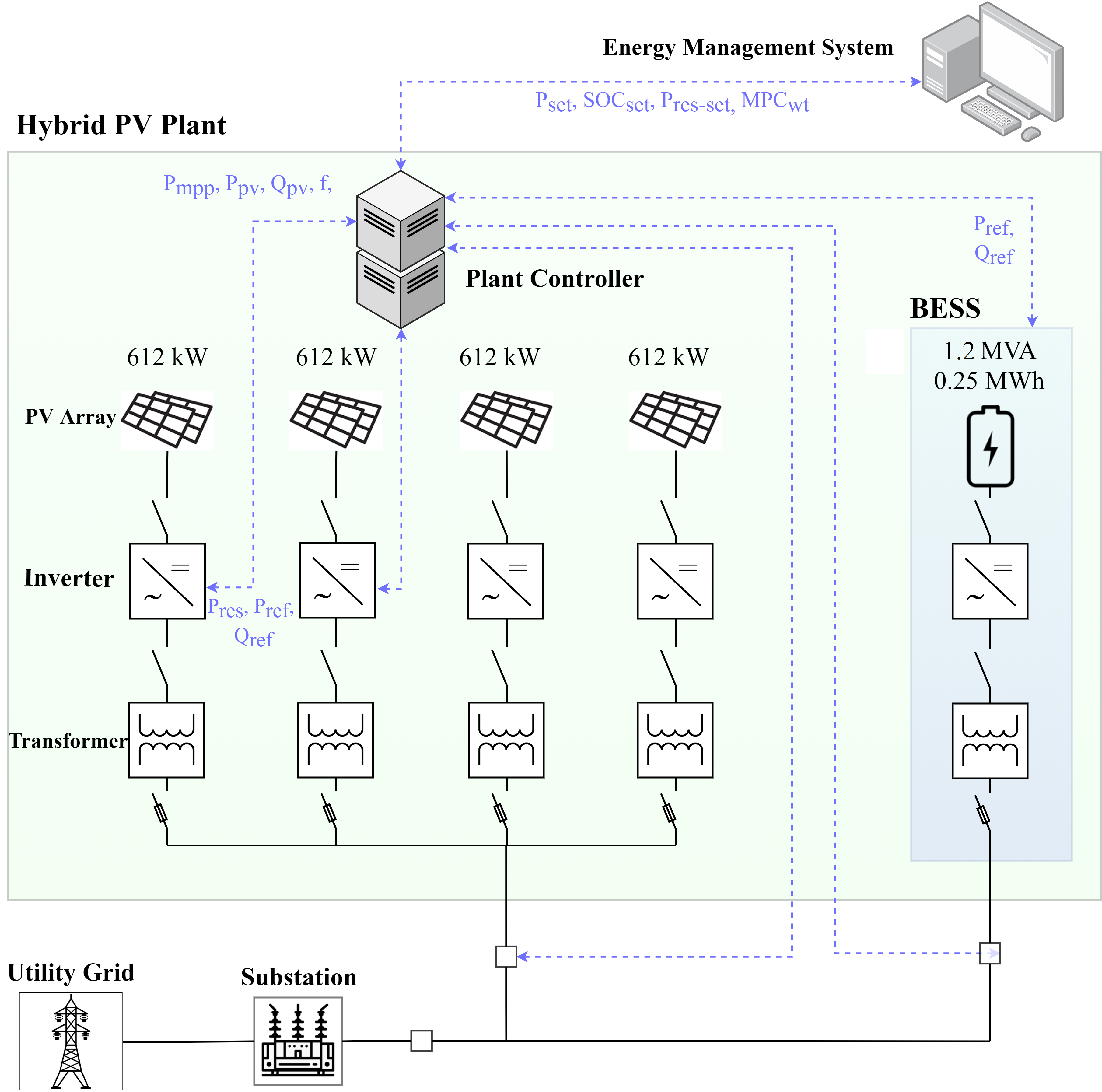}}
	\caption{Configuration of the hybrid PV plant.}
	\label{fig:hybrid_plant}
\end{figure}

To validate the proposed algorithm, a testbed is set up on an OPAL-RT real-time simulator. As shown in Fig. \ref{fig:hybrid_plant}, the hybrid PV plant consists of four equal 612 kWdc PV arrays, a plant controller, and a 1.2 MVA grid-following BESS. The BESS has a 1 MW/0.25 MWhdc lithium-ion battery unit and assists with the intra-minute power management. Table \ref{parameters_MPC} lists the simulation parameters. For more details on the MPC implementation, PV forecasting representation, and simulation settings, please refer to  \cite{paduani2022real}.

\begin{figure}[!htb]
	\centerline{\includegraphics[width=0.5\textwidth]{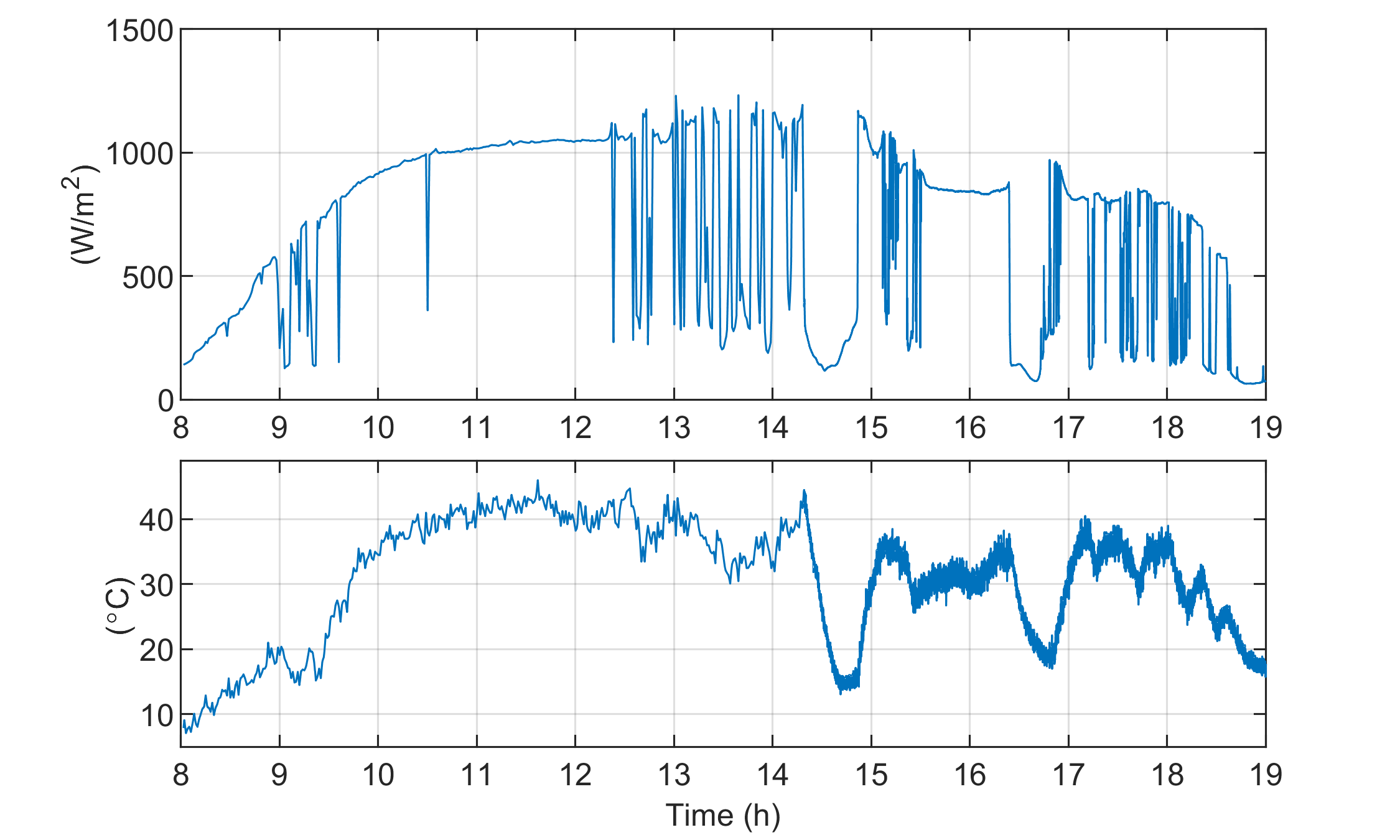}}
	\caption{Irradiance and temperature data from day 1.}
	\label{fig:irrad_day1}
\end{figure}

\begin{figure}[!htb]
	\centerline{\includegraphics[width=0.5\textwidth]{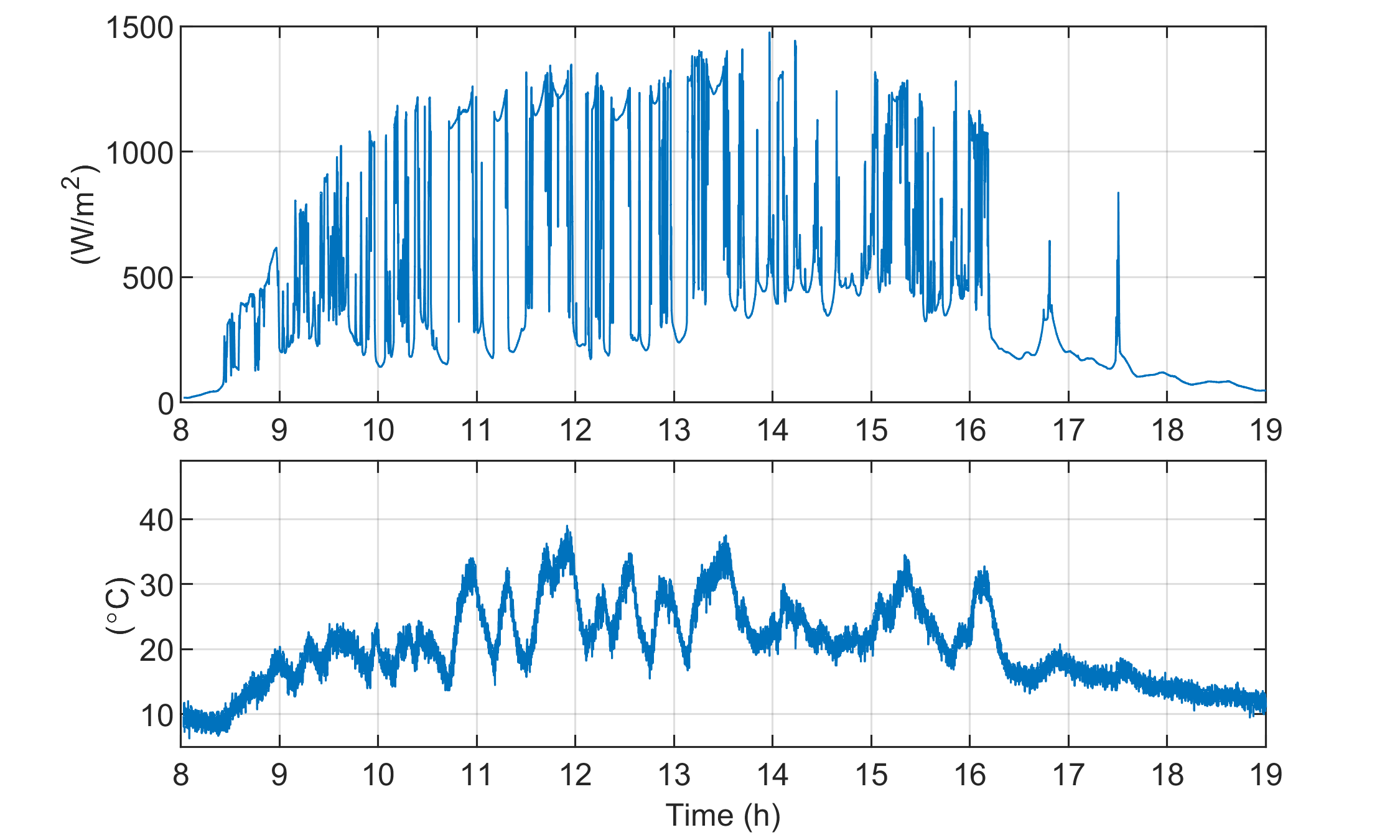}}
	\caption{Irradiance and temperature data from day 2.}
	\label{fig:irrad_day2}
\end{figure}

The high-resolution (one-second) irradiance and temperature data used to set up the simulation are collected by Strata Solar at a 5.04 MW solar farm located in North Carolina, USA, on April 8$^{\mathrm{th}}$ (Fig. \ref{fig:irrad_day1}) and 9$^{\mathrm{th}}$ (Fig. \ref{fig:irrad_day2}), 2022.

\renewcommand{\arraystretch}{1.1}
\begin{table}[htb]
	\caption{Hybrid PV Plant Model Parameters}
	\begin{center}
		\begin{tabular}{|>{\columncolor[gray]{0.85}} c|c|c|}
			\hline
			
			&Power&4$\times$\SI{612}{\kilo\watt}\\\cline{2-3}
			&Module&CS6P-250P\\\cline{2-3}
			&Size (parallel$\times$series) &153 $\times$ 16\\\cline{2-3}
			\multirow{-4}{*}{PV Arrays}&V$_{\mathrm{mp}}$, I$_{\mathrm{mp}}$& \SI{481.6}{\volt}, \SI{1270}{A}\\
			\hline\hline
			&Power, Frequency&\SI{500}{\kilo VA}, \SI{60}{\hertz} \\\cline{2-3}
			&Voltage (line-line) &200 V (RMS) \\\cline{2-3}
			&$\eta_{\mathrm{pv}}$ & 96.5\% \\\cline{2-3}
			&Ramp Rate & 0.2 p.u./s\\\cline{2-3}
			&L$_{\mathrm{f}}$, r$_{\mathrm{L}}$&\SI{100}{\micro\henry}, \SI{3}{\milli\ohm}\\\cline{2-3}
			&C$_{\mathrm{dc}}$& \SI{5000}{\micro\farad}\\\cline{2-3}
			&PI (v$_{\mathrm{dc}}$)& K$_{\mathrm{p}}$ = 1, K$_\mathrm{{i}}$ = 250\\\cline{2-3}
			\multirow{-8}{*}{PV Inverter}&PI (i$_{\mathrm{d}}$, i$_{\mathrm{q}}$)& K$_{\mathrm{p}}$ = 0.7, K$_{\mathrm{i}}$ = 50\\
			\hline\hline
			%&$\Delta G_{\mathrm{max}}$ & \SI{200}{\watt}/m$^{2}$/s \\\cline{2-3}
			%&$\Delta T_{\mathrm{max}}$ & \SI{3}{\degree C}/min \\\cline{2-3}
			%& Measurement window & 100 samples \\\cline{2-3}
			%&$a$ ($\eta$ gain) & 3 \\\cline{2-3}
			%&Damping $\eta$ range& [$10^{{\text-}6}, 10^{\text{-}3}$]\\\cline{2-3}
			%&Sampling freq. ($f_{\mathrm{s}} = 1/$T$_{\mathrm{s}}$)& %\SI{20}{\hertz}\\\cline{2-3}
			%\multirow{-7}{*}{\shortstack{Maximum\\ Power\\ Point\\ %Estimation}}&\shortstack{LM period ($T_{\mathrm{LM}}$)}& \SI{5}{\second}\\
			%\hline\hline
			%&Frequency ($f_{\mathrm{step}} = 1/$T$_{\mathrm{step}}$)& \SI{4}{\hertz} %\\\cline{2-3}
			%&$V_{\mathrm{step}\text{-}\mathrm{min}}$ / $V_{\mathrm{step}\text{-}\mathrm{b}}$ %& 0.75 / 2 V\\\cline{2-3}
			%& $K_{\mathrm{tr}}$  & 0.002\\\cline{2-3}
			%& Threshold $\Delta P_{\mathrm{ref,th}}$ & \SI{50}{k\watt/s} \\\cline{2-3}
			%&Threshold dp$_{\mathrm{th}}$ & \SI{15}{k\watt}\\\cline{2-3}
			%\multirow{-6}{*}{\shortstack{Flexible\\ Power\\ Point\\ Tracking}}&Threshold $dp/dv$& \SI{667}{\watt / \volt}\\
			
			\hline\hline
			&Power, Frequency & 1.2 MVA, \SI{60}{\hertz} \\\cline{2-3}
			&Voltage (line-line)& 480 V (RMS)\\\cline{2-3}
			&Ramp Rate & 0.2 p.u./s\\\cline{2-3}
			&$C_{f}$, $L_{f}$ & 2 mF, \SI{100}{\micro\henry}\\\cline{2-3}
			&PR controller &  K$_{\mathrm{p}}=2$, K$_{\mathrm{r}}=100$\\\cline{2-3}
			& $\eta_{\mathrm{charge}}$ (dc/ac) & 96.5\% \\\cline{2-3}		\multirow{-7}{*}{BESS}&$\eta_{\mathrm{discharge}}$ (ac/dc) & 96.5\%\\
			
			\hline\hline
			&Power, Capacity & 1 MW / 0.25 MWh \\\cline{2-3}
			&Voltage & 1600 V\\\cline{2-3}
			&Cells (series$\times$parallel) & 441$\times$9 \\\cline{2-3}
			&$C_{\mathrm{ts}}$, $R_{\mathrm{ts}}$ (per cell)& 440.57 F, 2 m$\Omega$ \\\cline{2-3}
			&$C_{\mathrm{tl}}$, $R_{\mathrm{tl}}$ (per cell)& 17111 F, 4.2 m$\Omega$ \\\cline{2-3}
			& $R_{\mathrm{s}}$ (per cell) & 1.3 m$\Omega$ \\\cline{2-3}		\multirow{-7}{*}{Battery}& $Q_{\mathrm{c}}$ (represents Ah) & 160 F\\
			\hline
		\end{tabular}
	\end{center}
	\label{parameters_MPC}
\end{table}

%Note that the irradiance and temperature measurements correspond to the incident irradiance on the PV panels (compensated), and the PV panels cell temperature. 

\subsection{Extended Kalman Filter Performance}
%\textcolor{red}{NL: I am confused here.  I think what you meant to compare is two extreme cases, where the error is biased, i.e., $+5\%$ and $-5\%$  with the baseline case, where the estimation has $\pm5\%$ error. Need to have a quick discussion here as the result discussion here is too short and readers will not understand why it is even necessary for you to compare the model estimation errors.}

\begin{figure}[!htb]
	\centerline{\includegraphics[width=0.5\textwidth]{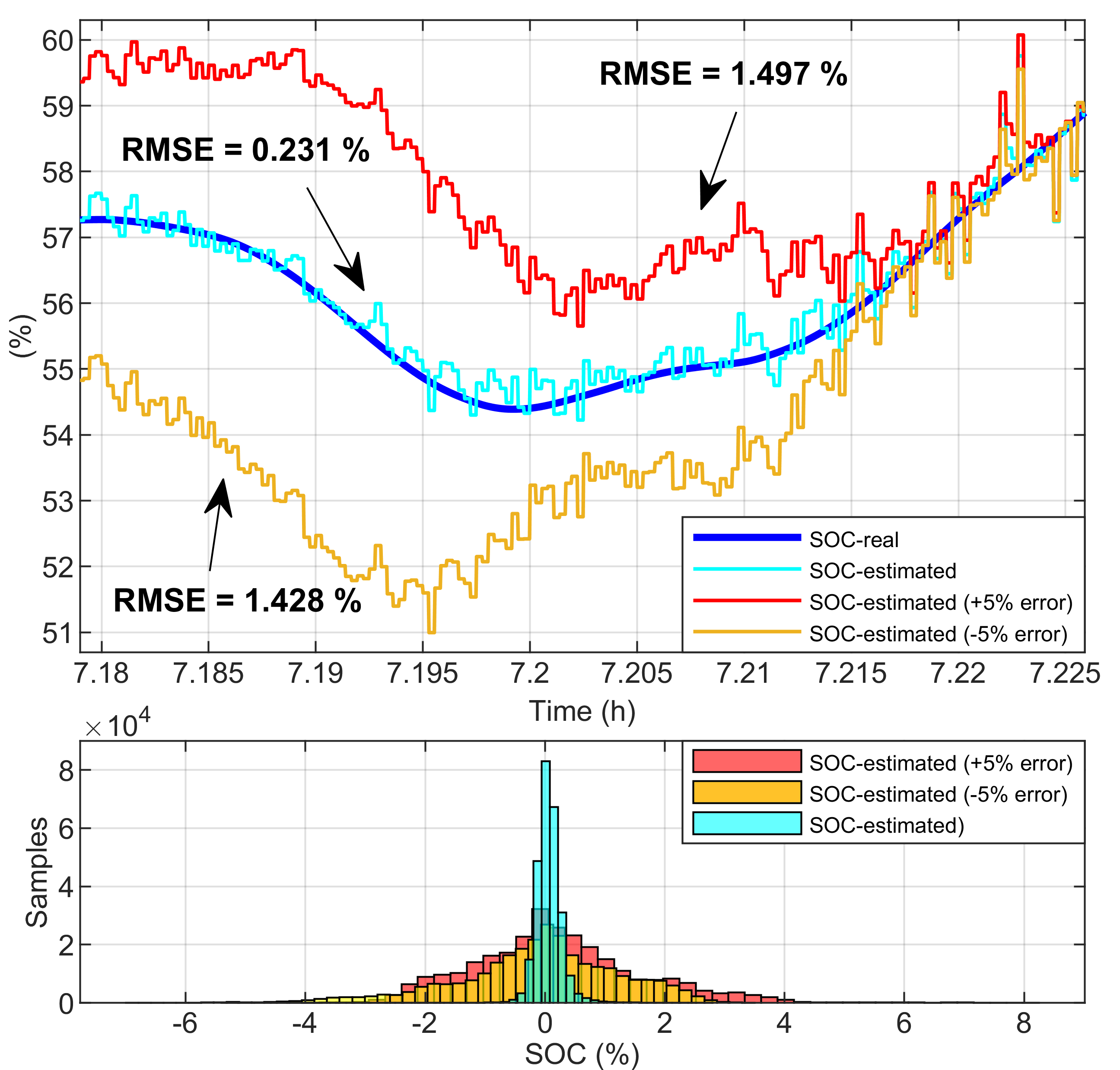}}
	\caption{Performance of the EKF SOC estimation considering parameter errors.}
	\label{fig:soc_kalman2}
\end{figure}

In this section, we conduct an error sensitivity analysis for the EKF-based SOC estimation to demonstrate its limitations. This is done by adding errors to the parameters of the battery mathematical model given by (\ref{eq:A_matrix}) to represent cases where the battery parameters cannot be accurately estimated.
%, representing a case in which the method utilized for extracting the battery parameters has inaccuracies. 
Three scenarios are studied: (i) all battery parameters, i.e. ($R_{\mathrm{s}}$, $R_{\mathrm{ts}}$, $R_{\mathrm{tl}}$, $Q_{\mathrm{c}}$, $C_{\mathrm{ts}}$, and $C_{\mathrm{tl}}$) are modeled with +5\% errors, (ii) all battery parameters are modeled with -5\% errors, and (iii) the base case where battery parameters are known. Results are displayed in Fig. \ref{fig:soc_kalman2}.

The RMSE errors in case (iii) are bounded by cases (i) and (ii). In (i), the SOC estimation was mostly above the actual SOC during discharging modes, and below the actual SOC during charging modes. The opposite was found for case (ii). In reality, a perfect extraction of the battery parameters is never possible because the battery parameters vary with respect to temperature, aging, etc. Therefore, the EKF estimation will be within an error margin. Consequently, in this paper, we set up a BESS SOC operation range to avoid overcharging.

\begin{figure}[!htb]
	\centerline{\includegraphics[width=0.5\textwidth]{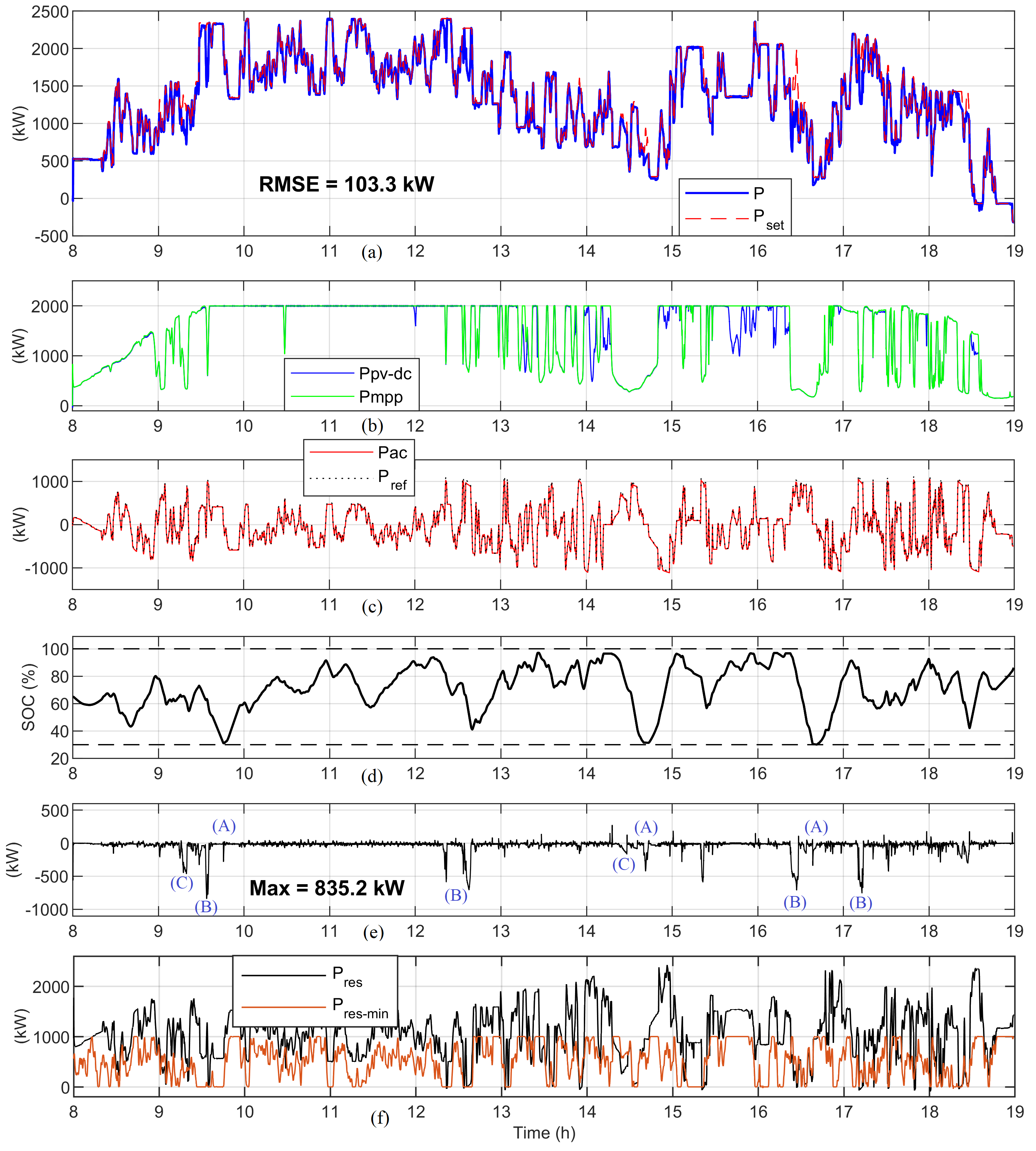}}
	\caption{Hybrid PV Plant performance in day 1 (case 1): (a) hybrid PV plant output power, (b) PV output power, (c) BESS output power, (d) battery SOC, and (e) power output errors, (f) power reserves.}
	\label{fig:ptrack_d1}
\end{figure} 

%From the results, it was found that neither the transient error spike caused by the change between battery charging or discharging modes, nor the error offset caused by including modeling errors inside the EKF battery SS model caused the EKF to lose its convergence.

\subsection{Power Regulation Performance}

This section analyzes the capability of the hybrid PV plant to follow power setpoints while operating under high irradiance intermittency conditions. To provide regulation services, the power setpoint of the hybrid PV plant is determined in two steps. First, the forecasted 30-minute average output power of the PV plant (assuming the forecasting error is within $\pm$10\%) is used to define the baseline power output of the plant for every 30-minute operation interval. Then, during each 30-minute interval, a 2-seconds 1 MW PJM regulation D signal (0.5 MW up/down with zero mean) is superimposed onto the 30-minute baseline to obtain the 2-second power setpoint.

In this case, the plant's power reserves setpoint ($P_{\mathrm{ref}}^{\mathrm{reserves}}$) starts as 500 kW, and is reduced down to 0 as the regulation signal ranges from 0 to positive 500 kW. This means that when providing upward regulation (power injection increase), the power reserves constraint is alleviated based on the power request increase. Alternatively, the plant could have been set to provide only downward regulation while continuously maintaining power reserves for FFR.

Results of day 1 (case 1) are presented in Fig. \ref{fig:ptrack_d1}. In this case, there are three main reasons for power output errors: (A) BESS SOC depletion, (B) power balance limitations due to extreme cloud coverage, and (C) power reserves constraints, which can be observed in Fig. \ref{fig:ptrack_d1}(f) when the actual power reserves (in black) becomes smaller than the minimum power reserves requested (in red). Each cause of error is respectively labeled in Fig. \ref{fig:ptrack_d1}(e).

Reason (C) can happen if high clouding events coincide with moments of high power setpoint request. When this happens, if the hybrid PV plant is asked to maintain power reserves, the problem becomes infeasible because either the power setpoint or the power reserves have to be violated. The trade-off between these two can be adjusted by increasing the softness of the power reserves constraint, where the operator can prioritize tracking power output setpoints or maintaining minimum power reserves for FFR services.

Errors originated from battery SOC depletion can be observed in Fig. \ref{fig:ptrack_d1}(e) around hours 9.75, 14.75, and 16.6, whereas errors due to the hybrid PV plant reaching its output power limits can be seen at hours 9.6, 12.6, and 17.25. At hour 17.25, for example, the PV availability drops below 500 kW. This is because although the battery operates at the nominal output power (1 MW), the power setpoint of 2 MW cannot be maintained, causing a large output error. Errors due to power reserves reaching their limits can be seen around hours 9.3, 14.4 and 16.4 (see Fig. \ref{fig:ptrack_d1}(f)).

\begin{figure}[!t]
	\centerline{\includegraphics[width=0.5\textwidth]{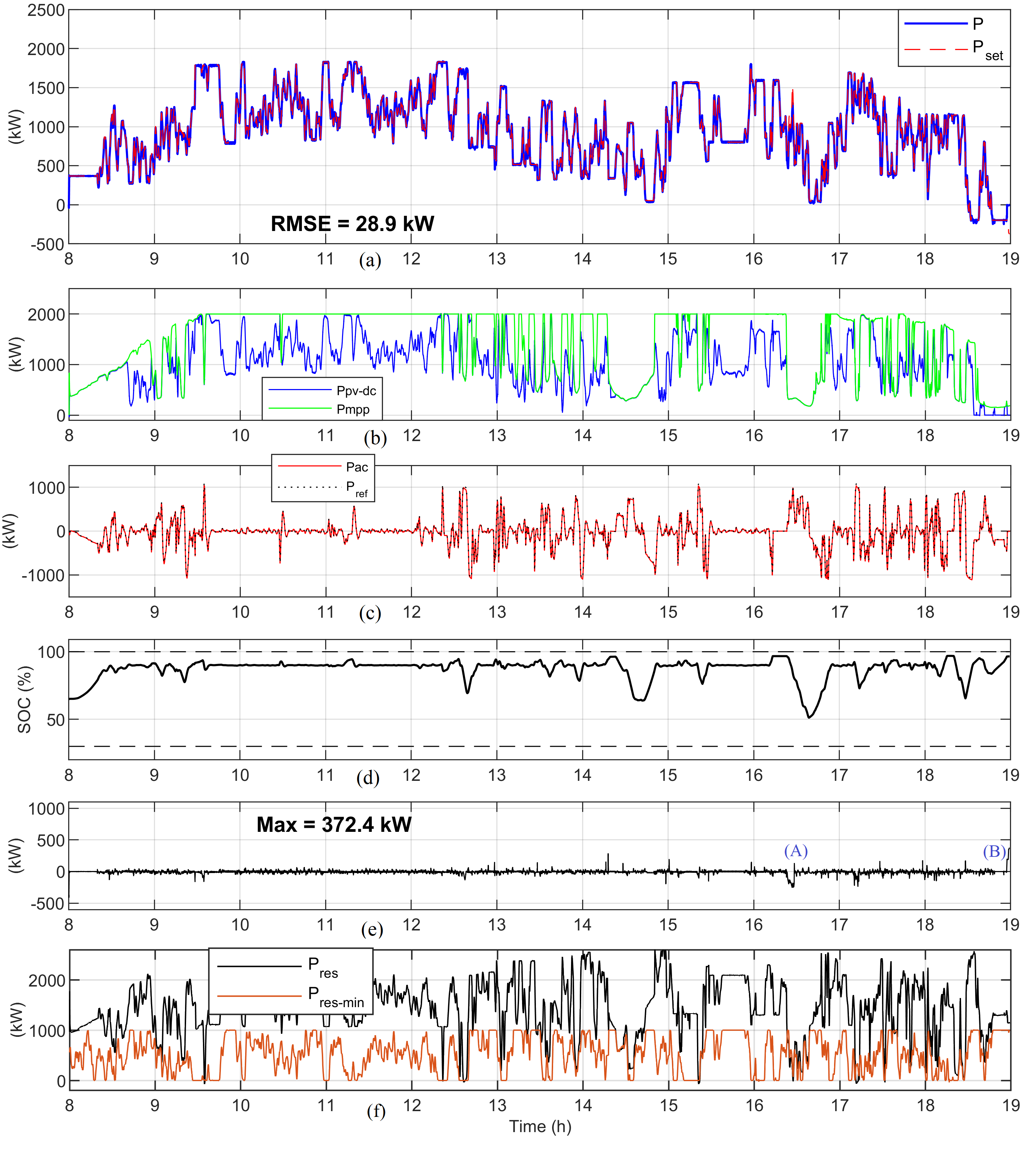}}
	\caption{Hybrid PV Plant performance in day 1 (case 2): (a) hybrid PV plant output power, (b) PV output power, (c) BESS output power, (d) battery SOC, and (e) power output errors, (f) power reserves.}
	\label{fig:ptrack_d1_low}
\end{figure} 

One solution to reduce the power setpoint error is to reduce the average output power requested such that the expected power from the plant assumes a more conservative approach. Thus, in Case 2, we reduce the average requested power by 25\%. As shown in Fig. \ref{fig:ptrack_d1_low}, the plant can provide a much closer power setpoint tracking if a less aggressive request is demanded from it. In this case, the plant power output RMSE reduced from 103.3 to 28.9 kW, close to four times smaller. Furthermore, by comparing Fig. \ref{fig:ptrack_d1}(f) with \ref{fig:ptrack_d1_low}(f), it can be noticed how in the second case the system is able to maintain higher power reserves. In this case, there are two main sources of errors: (A) power reserves limitations, and (B) BESS reaching upper SOC limits. Note the BESS upper SOC is limited to a value below 100\% for two reasons: (i) providing a safety margin to accommodate for SOC estimation errors, and (ii) maintaining enough headroom for providing FFR services at anytime. 

\begin{figure}[!t]
	\centerline{\includegraphics[width=0.5\textwidth]{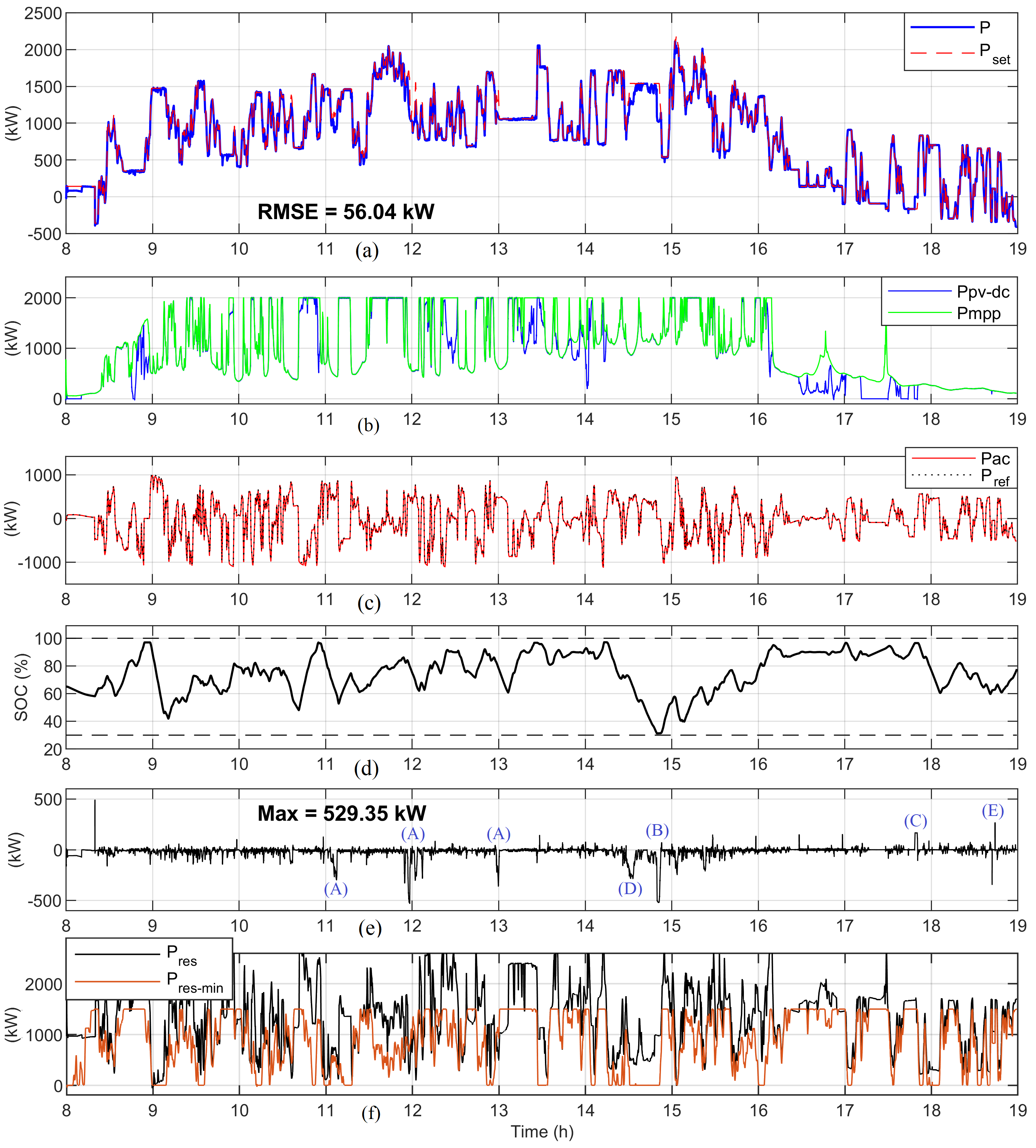}}
	\caption{Hybrid PV Plant performance in day 2: (a) hybrid PV plant output power, (b) PV output power, (c) BESS output power, (d) battery SOC, and (e) power output errors, (f) power reserves, (f) power reserves.}
	\label{fig:ptrack_d2}
\end{figure} 

Figure \ref{fig:ptrack_d2} displays the performance for the second day, corresponding to an extreme scenario of very high irradiance intermittency. Note there is a spike in the power output error around hour 8.3. This is because the regulation signal is activated at hour 8.3 (after 20 minutes). Thus, if a very high regulation value is received right upon activation (see Fig. \ref{fig:ptrack_d2}(f), a large instantaneous error will occur because the plant must ramp up to its new setpoint.

In this case, the causes of errors are: (A) power reserves limits, (B) BESS SOC depletion, (C) BESS reaching upper SOC limits, (D) SOC usage optimization, (E) ramping limits. Note the error at hour 14.5 occurs even though the battery could have increased its power output. This is because after considering (i) the available PV (and its forecast), (ii) the requested power output, and (iii) the BESS increased losses at higher power outputs, the MPC optimization starts to degrade its power tracking performance to optimize the usage of available BESS SOC. On the other hand, during hours 11.1, 12, and 13, there are errors even though there is sufficient battery SOC, and the battery is not reaching its nominal ratings. Those errors occur because the system reaches its minimum power reserves constraint, and thus constraints' violation weights are considered to decide how much of the power output should be sacrificed in exchange for reserve violations. This relation can be tuned as needed by adjusting the minimum and maximum output variables ECR gains ($y_{i,\mathrm{ECR}}^{\mathrm{min}}$, $y_{i,\mathrm{ECR}}^{\mathrm{max}}$), given in the Appendix section.

%\begin{figure}[!htb]
%	\centerline{\includegraphics[width=0.8\textwidth]{Chapter-5/figs5/Preserves_violation2.png}}
%	\caption{Moment of power reserves violations (a) PV and BESS output power, (b) power reserves and regulation D signal.}
%	\label{fig:Pres_violation}
%\end{figure} 

In addition, a positive error is observed around hour 17.8. It occurred because the battery reached its upper SOC limit, so the hybrid PV plant was incapable of absorbing power for a downward regulation. %Note that in this work the actual battery maximum SOC is set to 97.5\% (given in the appendix), to account for errors in the EKF estimation, and a hard threshold is set at SOC 96\% so that whenever the battery SOC reaches 96\%, its maximum current constraint its updated to zero with a ramp rate limiter of 50 A per second, such that until the SOC drops, the battery cannot be charged further. Once the charge is below 96\%, the constraint is once again ramped back to the original setting (-550 A, in this work). The hard threshold can slightly reduce the MPC optimization performance, since it will expect the BESS SOC to reach up to 97.5\%, but it is important to avoid numerical instabilities in the solver in case a small violation in the battery SOC occurs.\footnote{Recall that the battery SOC constraints are designed with a small degree of softness.}

\subsection{Comparison to a Thermal Machine}

Next, the power regulation of the hybrid PV plant is compared to that of a typical PJM thermal unit with a ramping of 0.8 MW per minute \cite{kirby2005method} during day 2. The machine model is built with governor, turbine, and reheater time constants given in \cite{abd2018firefly}. The hybrid plant SOC setpoint is dropped from 90\% to 80\% to leave more room for downward regulation, and the average power requested is reduced by 20\%. The same power setpoint is requested from both systems, with a power regulation signal of 1.5 MW. Because the thermal unit cannot absorb power, its power setpoint is offset by 750 kW for the entire operation. 

\begin{figure}[!htb]
	\centerline{\includegraphics[width=0.5\textwidth]{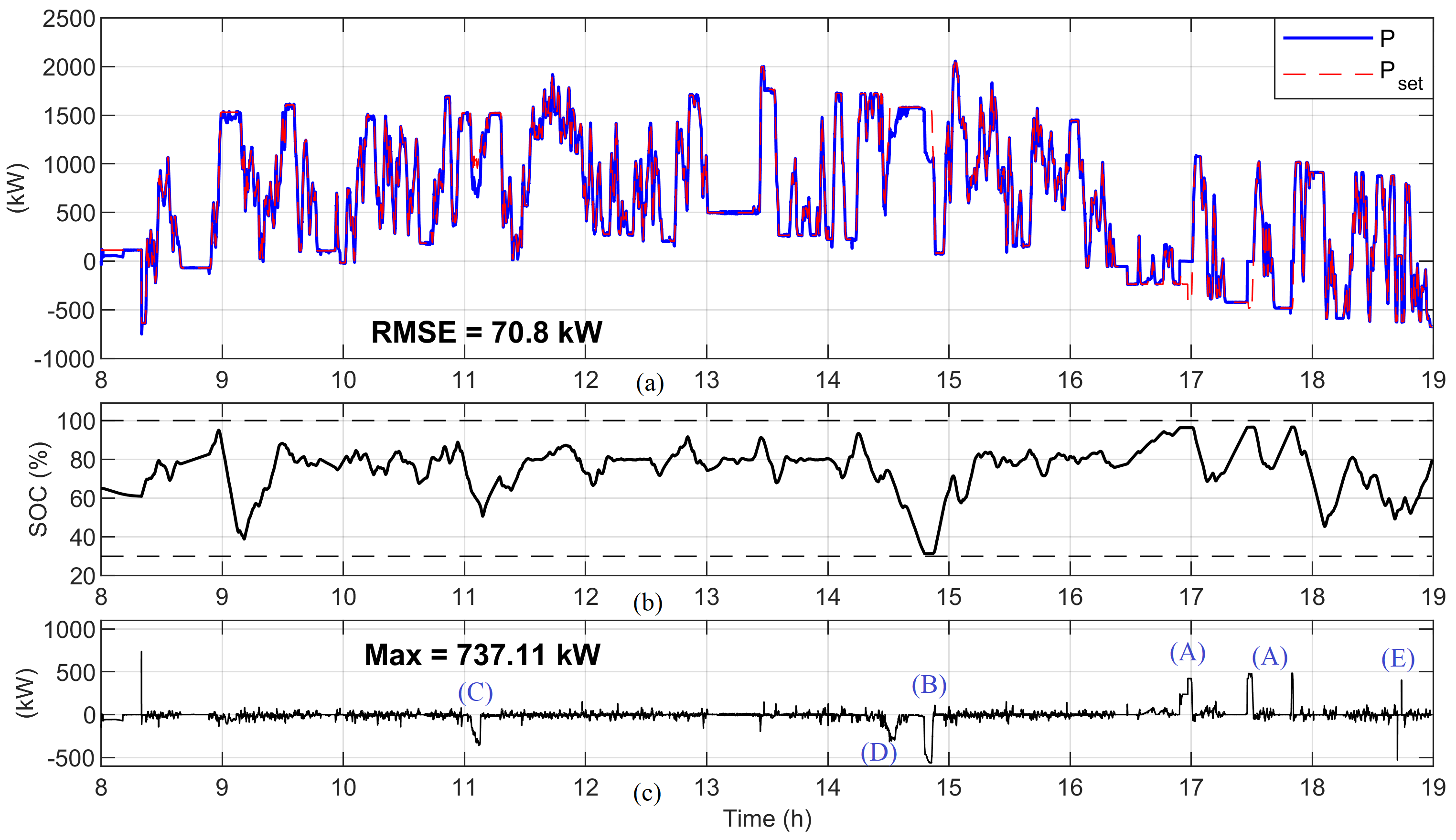}}
	\caption{Hybrid PV Plant performance in day 2 (providing 1.5 MW regulation service): (a) power output, (b) battery SOC, (c) power output error.}
	\label{fig:Ptrack_d2_dg}
\end{figure}

\begin{figure}[!htb]
	\centerline{\includegraphics[width=0.5\textwidth]{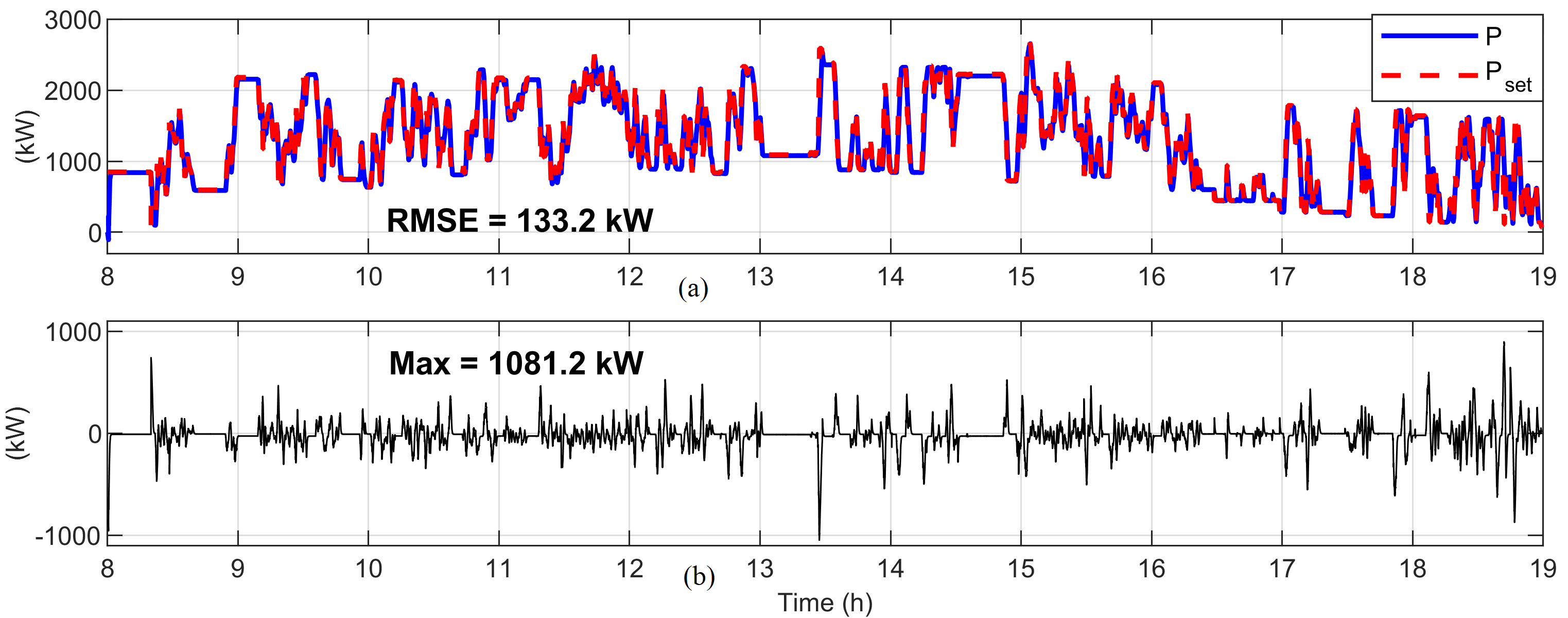}}
	\caption{Thermal generator performance following same power setpoint in day 2 (providing 1.5 MW regulation service): (a) power output, (b) power output error.}
	\label{fig:Ptrack_d2_750}
\end{figure}

As shown in Figs. \ref{fig:Ptrack_d2_dg} and \ref{fig:Ptrack_d2_750}, compared to the hybrid PV plant, the thermal machine has a slower response rate and consequently, a significant higher power output error. In Fig. \ref{fig:Ptrack_d2_dg}(c), we highlight five factors causing power output error in the hybrid PV plant operation: (A) hitting battery upper SOC limit, (B) hitting battery lower SOC limit, (C) reaching power reserves limits, (D) limited by the SOC usage optimization requirement, and (E) hitting ramping constraints. 

\begin{figure}[!htb]
	\centerline{\includegraphics[width=0.5\textwidth]{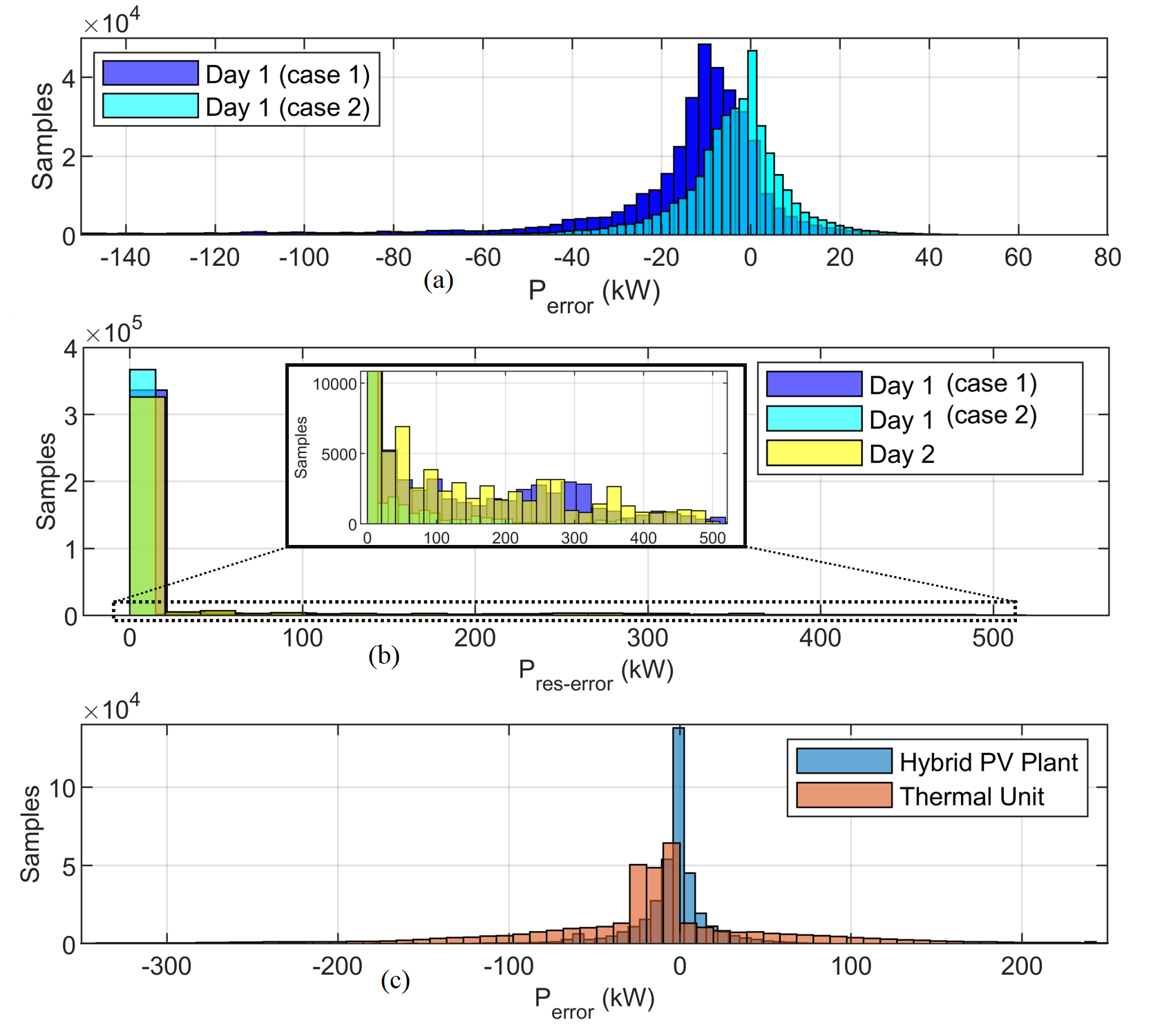}}
	\caption{Comparison of setpoint tracking errors: (a) power output of day 1 (cases 1 and 2), (b) power reserves of day 1 (cases 1 and 2) and day 2, (c) power output between hybrid PV plant and thermal unit in day 2.}
	\label{fig:histograms_combined}
\end{figure}

Figure \ref{fig:histograms_combined} presents an analysis of the errors in power output and power reserves requested through the presented cases. As can be seen in Figs. \ref{fig:histograms_combined}(a) and \ref{fig:histograms_combined}(b), when reducing the average power output request by 25\% from case 1, the performance tracking performance can be significantly improved, with power reserves being successfully maintained throughout 96.65\% of the operation time. This demonstrates that an oversized hybrid PV plant can become a reliable source of power reserves even if the plant follows variable power setpoints in days of high irradiance intermittency. 

Figure \ref{fig:histograms_combined}(c) compares the power output error between the Hybrid PV plant and the thermal unit. The results show the hybrid PV plant can provide superior regulation performance due to its faster IBR controls. Note that the thermal unit has a typical ramping of 0.8 MW per minute. In this test, we set the hybrid PVplant ramping at 0.6 MW per second, although in practice, the PV plant ramping could be even faster based on the agreement with the system operator. Moreover, it is important to highlight that day 2 is a day of very high intermittency (Fig. \ref{fig:irrad_day2}), selected to demonstrate performance limitations for the worst case scenario.

 %Results are shown in Fig. \ref{fig:Ptrack_d2_750}. It can be seen that the hybrid PV plant can present a smoother tracking while there is enough SOC available to maintain the regulation. However, due to the lower power setpoint requested around hours 17 to 18, the battery reached its upper SOC limit, being then unable to provide downward regulation. 

\section{Conclusion}

This work proposes a control strategy to optimally operate a hybrid PV plant while considering power reserves for providing FFR services. The system combines optimal control with the power curtailment algorithm introduced in \cite{paduani2021unified} to form a PV plant capable of maintaing reserves for providing both regulation and/or FFR services for grid support. An adaptive MPC is utilized to handle the nonlinearities of the plant model, and detailed EMT models are developed to provide an accurate representation of components losses and limitations. Results demonstrate the system can provide high quality power regulation services as well as maintain robust power reserves without the need for large BESS storage. Furthermore, the proposed framework is highly customizable, such that constraint softness and adjustable control objective weights can be used to shift priorities between power regulation, power reserves, or battery SOC management in real-time.

\section*{Acknowledgment}
The authors thank Abhishek Komandur,  Taylor Adcox, and Laura Kraus with Strata Solar for their technical support and provision of utility-scale solar farm datasets.

\appendix[]

 $T_{\mathrm{mpc}}$ = 3 s, p = 400, m = 20, $a_{1} = 8.4073$, $a_{2} = -19.892$, $a_{3} = 11.497$, $a_{4} = 4.161$, $a_{5} = -4.5533$, $a_{6} = 0.34365$, $a_{7} = 0.64685$, $a_{8} = 3.5016$, $w_{1}^{y} = 3$, $w_{2}^{y} = 0$, $w_{3}^{y} = 0.1$, $w_{4}^{y} = 0$, $w_{5}^{y} = 0$, $s_{1}^{y} = 2$ MVA, $s_{1}^{y} = 1200$ A, $s_{3}^{y} = 1$, $s_{4}^{y} = 4$ MVA, $s_{5}^{y} = 2$ MVA, $w_{1}^{\Delta u} = 0.01$, $w_{2}^{\Delta u} = 0.002$, $s_{1}^{\Delta u} = 280$ A/s, $s_{2}^{\Delta u} = 800$ kVA/s, $y_{1,\mathrm{ECR}}^{\mathrm{min}} = 1$, $y_{1,\mathrm{ECR}}^{\mathrm{max}} = 1$, $y_{2,\mathrm{ECR}}^{\mathrm{min}} = 0.5$, $y_{2,\mathrm{ECR}}^{\mathrm{max}} = 0.5$, $y_{3,\mathrm{ECR}}^{\mathrm{min}} = 0.5$, $y_{3,\mathrm{ECR}}^{\mathrm{max}} = 0.5$, $y_{4,\mathrm{ECR}}^{\mathrm{min}} = 4$, $y_{4,\mathrm{ECR}}^{\mathrm{max}} = 5$, $y_{5,\mathrm{ECR}}^{\mathrm{min}} = 0.3$, $y_{5,\mathrm{ECR}}^{\mathrm{max}} = 0.3$, $u_{1}^{\mathrm{min}} = -130$ A, $u_{1}^{\mathrm{max}} = 130$ A, $u_{2}^{\mathrm{min}} = -400$ kVA, $u_{2}^{\mathrm{max}} = 400$ kVA, $y_{1}^{\mathrm{min}} = -1$ MVA, $y_{1}^{\mathrm{max}} = 3$ MVA, $y_{2}^{\mathrm{min}} = -550$ A, $y_{2}^{\mathrm{max}} = 650$ A, $y_{3}^{\mathrm{min}} = 0.295$, $y_{3}^{\mathrm{max}} = 0.975$, $y_{4}^{\mathrm{min}} = -3$ MVA, $y_{4}^{\mathrm{max}} = 3$ MVA, $y_{5}^{\mathrm{min}} = 0$, $y_{5}^{\mathrm{max}} = 2$ MVA.

%\ifCLASSOPTIONcaptionsoff
%  \newpage
%\fi
\bibliographystyle{IEEEtran}
\bibliography{OptimalHybridPlant}

\end{document}